\documentclass[12pt, a4paper]{article}

\usepackage[utf8]{inputenc}
\usepackage{longtable}
\usepackage{graphicx}

\usepackage[square,numbers]{natbib}
\usepackage{ulem}

\usepackage{amsmath,amsfonts, amsthm, amssymb}
\usepackage{newlfont}
\usepackage{fancyhdr}
\usepackage[Glenn]{fncychap}
\usepackage{fullpage}

\usepackage[T1]{fontenc}
\usepackage{array,multicol}
\usepackage{setspace} % Espaciado entre lineas
\def\R{\mat{R}}

\def\Kk{{\cal K}}

\def\DD{{\cal D}}

\newcommand{\Db}{{\boldsymbol D}}

\def\K{{\cal K}}

\newcommand{\nnabla}{{\boldsymbol \nabla}}
\newcommand{\Fb}{{\boldsymbol F}}
\newcommand{\Lb}{{\boldsymbol L}}

\newcommand{\bs}{{\boldsymbol \sigma}}
\newcommand{\bbs}{{\boldsymbol b}}
\newcommand{\F}{{\cal F}}
\newcommand{\bm}{{\boldsymbol m}}
\newcommand{\bu}{{\boldsymbol v}}
\newcommand{\D}{{\boldsymbol D}}
\newcommand{\W}{{\boldsymbol W}}
\newcommand{\bW}{\boldsymbol  W}

\newcommand{\bFs}{{\boldsymbol {\cal F}}}
\newcommand{\Rb}{{\boldsymbol R}}
\newcommand{\M}{{\boldsymbol M}}
\newcommand{\Pb}{{\boldsymbol P}}

\newcommand{\Sb}{{\boldsymbol S}}
\newcommand{\Ub}{{\boldsymbol U}}
\newcommand{\Cb}{{\boldsymbol C}}
\newcommand{\Zb}{{\boldsymbol Z}}
\newcommand{\Ib}{{\boldsymbol I}}

\newcommand{\Qb}{{\boldsymbol Q}}
\newcommand{\bV}{{\boldsymbol V}}
\newcommand{\bA}{{\boldsymbol A}}
\newcommand{\ba}{{\boldsymbol a}}

\newcommand{\bth}{{\boldsymbol \theta}}
\newcommand{\bOm}{{\boldsymbol \Omega}}
\newcommand{\bom}{{\boldsymbol \omega}}
\newcommand{\I}{{\boldsymbol I}}

\newcommand{\be}{{\boldsymbol e}}
\newcommand{\br}{{\boldsymbol r}}

\newcommand{\bw}{{\boldsymbol w}}
\newcommand{\bn}{{\boldsymbol n}}
\newcommand{\bc}{{\boldsymbol c}}
\newcommand{\bd}{{\boldsymbol d}}
\newcommand{\bq}{{\boldsymbol q}}

\newcommand{\ff}{{\boldsymbol f}}

\def\R{\mathbb{R}}

\numberwithin{equation}{section}

\usepackage{hyperref}
\hypersetup{
  colorlinks   = true,    % Colours links instead of ugly boxes
  urlcolor     = blue,    % Colour for external hyperlinks
  linkcolor    = blue,    % Colour of internal links
  citecolor    = blue     % Colour of citations
}

\begin{document}
%\english
% Large plastic deformation of   crystals: orientation attractors  in velocity gradient  driven processes
\title{Prediction of ideal orientations in velocity gradient-driven processes for large plastic deformations of crystals
%Orientation attractors  in velocity gradient  driven processes for large plastic deformations of crystals
}
\author{
	Jalal Smiri\footnote{LSPM, CNRS UPR3407, Université Sorbonne-Paris-Nord, Villetaneuse, France} 
	\footnote{now at  FEMTO-ST, Université Marie et Louis Pasteur, 25000-Besançon, France}, 
	O\u{g}uz Umut Salman$^*$, 
	and Ioan R. Ionescu$^{*}$\footnote{IMAR, Romanian Academy, 10587-Bucharest, Romania} \footnote{corresponding author, ioan.r.ionescu@gmail.com}
}\date{version : \today}

\maketitle

\begin{abstract}
We focus on the crystal lattice ideal orientations, also referred to as preferred or attractor orientations, in crystalline materials, and how they can be used to predict the final texture of polycrystals after manufacturing processes.
The simplified crystal plasticity (CP) models used here capture the main features of microstructural evolution in monocrystalline and polycrystalline materials undergoing velocity-gradient-driven processes, without considering hardening or softening effects. The evolution of the lattice orientation is described by a nonlinear ordinary differential equation, and a linear stability analysis is performed to identify the permanent orientations that act as attractors (i.e., the ideal or preferred orientations).
Although our linear stability analysis is generally applicable, it is detailed using a simplified two-dimensional model with three slip systems. This approach successfully predicts lattice orientation attractors for large strains by analyzing the interplay between deformation and rotation, initial orientation, and the interaction between different slip systems under applied loads.
Three fundamental problems in CP illustrate the effectiveness of the theory: polycrystal deformation under homogeneous velocity gradient loading, void evolution under radial loading, and slip band formation in a monocrystal.
High-resolution CP numerical simulations, enhanced using re-meshing techniques, provide further validation of our findings concerning the impact of initial crystallographic orientations, deformation mechanisms, and loading conditions on the evolution of orientation attractors and the final crystal texture.
\end{abstract}
%\newpage 
%\tableofcontents{}
%\newpage

\section {Introduction}

Metallic materials used in structural applications are commonly produced as polycrystals, composed of numerous grains, each with its own unique crystallographic orientation. This results in varying degrees of plastic deformation among grains, influenced by their orientation, geometry, interactions with neighbors, and applied loading \cite{HU201986}. Beyond this intergranular deformation heterogeneity, intragranular heterogeneity also emerges during plastic deformation, manifesting as localized, continuous, or discontinuous orientation spreads within individual grains \cite{SEDIGHIANI2022118167}.

The stability of crystallographic orientations plays a pivotal role in the evolution of material textures, particularly during plastic deformation. A deep understanding of this stability is essential for characterizing how materials respond to deformation in both mono- and polycrystalline structures. Accurately predicting crystallographic texture evolution is crucial for optimizing material properties, especially under conditions of large plastic strain and deformation localization. This evolution is primarily governed by lattice rotation, a kinematic consequence of slip and twinning, with the resulting microtexture determined by orientation stability.

Lattice rotation has been examined in various microstructural phenomena, such as the formation of fragmentation bands during cold rolling~\cite{Song2024-ag}, the dynamic texture evolution observed during shock compression~\cite{Avraam2023-yj}, the texture evolution steel subjected to tensile deformation \cite{cryst14020186}, and how orientation-dependent lattice rotation drives nucleation and growth of dynamic recrystallization features relevant to fragmentation of grains under large strain \cite{Chen2023-qe}. Analyzing these large, anisotropic deformations requires robust computational tools, including adaptive remeshing techniques~\cite{sartorti2024remeshing}, as well as theoretically rigorous modeling frameworks that effectively handle finite-deformation kinematics~\cite{Sen2022-bn, Rubin2024-ol}. Therefore, conducting a rigorous theoretical stability analysis of lattice rotation—aimed at identifying stable orientation attractors under extreme plastic strain—is a critical step. Such analysis is essential not only for validating complex finite-deformation models but also for fundamentally improving the predictability of texture evolution in highly deformed materials and localized shear zones.

During deformation, lattice orientations tend to stabilize around specific configurations, often referred to as lattice orientation attractors, or preferred or ideal orientations, which are crucial for predicting material behavior under stress. Texture evolution during deformation generates intragranular substructures with orientations that deviate from the parent grain's initial orientation. During continued deformation, these substructures evolve toward preferred orientations shaped by the imposed deformation mode and initial orientation, which subsequently influences the material's mechanical properties and overall texture \cite{SEDIGHIANI2022118167, ardeljan2015study, suwas2019texture, ASK2018167, zhou2024investigation,TANDOGAN2026106325,Petryk2023-ub}.

Crystallographic orientation is a key determinant of whether a grain will undergo uniform (homogeneous) or localized (heterogeneous) deformation. Stable orientations typically lead to more uniform deformation, whereas unstable orientations are often linked to strain localization and deformation inhomogeneity. Thus, the stability of crystallographic orientations becomes central to understanding texture development during plastic deformation.

The first steps in the  investigation of  texture evolution and preferred orientations were carried out  in \cite{PJD93,TBST08} for a simplified planar polycrystal with two systems  where  the partition of shearing among the slip systems is kinematically determined.   Other  
 papers  provide insight into how initial crystallographic orientations affect the stability of lattice configurations during deformation  for a specific loading such as uni-axial \cite{GTFC19, SLB24}, equibiaxial tension  \cite{hajian2015deformation, khajeh2016ideal, hajian2016prediction, li2008orientation}, equal-channel angular extrusion channel \cite{li2008orientation}, torsion \cite{ABBL22} or  rolling \cite{kestens2016texture, chen2020oriented}.  Typically, these simulations use an Eulerian framework with zero rotation rate
 to isolate the effects of external forces on orientation evolution.  Significant effort was also made  for modeling texture evolution due to dynamic recrystallization of ice  to deduce the existence of  orientation attractors that maximize the resolved shear stress on the easiest slip system in the crystal, see \cite{CRMECA_2024__352_G1_99_0, 2015FrEaS381M, CHAUVE2015116, chauve2017investigation}.

%Conventionally, large deformation simulations are performed in a Lagrangian framework, where the mesh follows the deforming material. While effective for moderate strains, as deformation progresses, mesh distortion introduces significant errors and ultimately leads to simulation failure. To address these challenges, this paper adopts an Eulerian approach, which is particularly effective for simulating extreme deformations. In the Eulerian formulation, a fixed grid allows materials to flow through without distorting the mesh, making it ideal for analyzing complex material behavior during severe plastic deformation.

The CP  elasto-(visco)plastic model requires a consistent interplay between Eulerian
%(better suited for plasticity)
 and Lagrangian
 % (better suited for elasticity) 
  descriptions. However,  in applications involving large deformations of metals the elastic component of the deformation is small  with respect to the inelastic one,  and can be neglected  by using  a rigid-(visco)plastic approach (see for instance  \cite{lebtome93,kok02}). The rigid-(visco)plastic assumption (neglecting elastic effects) may
be overly restrictive for some applications where elastic and plastic deformations are of the same order but it is well accepted for very large plastic deformations. 
This simplified model,  used here and   introduced by  \cite{Cazacu2010-hi}, takes important theoretical and numerical  advantages  by using only the Eulerian configuration.

The aim of this paper is to investigate the stability of lattice orientation in velocity gradient-driven processes for very large deformations. Our objective is to provide a general theoretical  framework for the stability analysis of crystal lattices that can model the evolution towards stable (ideal or preferred) orientations. These   ideal or preferred orientations depend on  the interplay between the imposed deformation and rotation and on the initial lattice orientation (initial crystallographic texture). To simplify the analysis, we consider only perfectly (visco-)plastic models without hardening or softening.  For velocity gradient-driven processes, the evolution of the lattice orientation is written as a nonlinear ordinary differential system. A linear stability analysis is then used to determine the permanent orientations that act as attractors, i.e., the ideal or preferred orientations. It should be noted that, even when an Eulerian framework is used, the stability analysis is material (i.e., it concerns the crystal orientation of a particle).

Since the analytical stability analysis of a 3D crystal is too complex, a more detailed investigation was conducted for a simplified 2D model with only three slip systems. This model is simple enough to allow analytical treatment yet rich enough to
capture the essential features of a 3D crystal.  We characterize the stable stationary orientations and deduce their basins of attraction from the interplay between the deformation and rotation of a velocity gradient-driven process. These basins of attraction are essential in predicting the final texture of a poly/monocrystal. 

Let us outline the paper's content. In Section 2, we recall from \cite{CI09}  the 3D rigid-(visco)plastic crystal model   and we propose 
a  stability analysis framework for lattice orientation in velocity gradient-driven processes. Since the general 3D case is too complex for an  analytic or numerical treatment  Section 3 provides detailed stability analysis for the simplified 2D model with three slip systems, characterizing ideal orientations and their attraction basins.    Section 4 presents three fundamental problems in CP, which illustrate the above stability analysis. The problems are as follows: (1) a polycrystal with a homogeneous velocity gradient and heterogeneous initial orientations, (2) a monocrystal with void growth under radial loading, and (3) slip band formation in a single crystal under uniaxial loading.  The theoretical results obtained for specific velocity-gradient-driven processes are compared to the numerical results of the associated boundary-value problems in all these cases.

\section{Analysis  of  velocity gradient  driven processes}

\subsection{ Eulerian rigid-plastic approach of  crystal  plasticity} \label{Chapter:model}

Consider a single crystal at time $t=0$,  free of any surface tractions and body forces and let choose this configuration, say  $\Kk_0 \subset \R^3$  as reference configuration of the crystal. Let $\Kk=\Kk(t)\subset \R^3$ denote the current configuration. The incorporation of lattice features is achieved through a multiplicative decomposition of the total deformation gradient $\Fb$ into elastic and plastic components:
\begin{equation}
	\Fb = \Fb^{e} \Pb.
	\label{eq:1}
\end{equation}
This decomposition implies a two-stage deformation process. First, $ \Pb$ transforms the initial reference state $\Kk_0$ to an intermediate state $\tilde{\Kk}$, characterized by plastic deformation only with no change in volume.   $\Pb$ is called the (visco)plastic deformation with respect to the reference configuration of a material neighborhood of the material point $X$ at time $t$. Then, $\Fb^{e}$ brings the body to the final configuration $\K$ through elastic deformation and rigid lattice rotation, i.e., $\Fb^e=\Rb\Ub^e$ where $\Rb$ denotes the rotation of the crystal axes with respect to its isoclinic orientation. 

%A constitutive law for the elastic part of the deformation (see \cite{McHugh2004-kl} for example) is needed. Simplest formulations assume a linear relationship between the second Piola-Kirchhoff stress tensor and the Right Cauchy-Green strain tensor  in the intermediate configuration $\tilde{\Kk}$, though higher order elastic moduli have also been considered \cite{Teodosiu2013-zq}. These incremental equations, relating total (plastic and elastic) stress to total deformation, can be equivalently mapped to the deformed configuration as done in \cite{Asaro1985-nm, Asaro1983-cw}. The original formulation in the undeformed configuration is summarized in \cite{Harren1988-gi,McHugh1993-mp,Cuitino1993-rp}.

Following \cite{Asaro1977-nn}, $\Pb$ is assumed to leave the underlying lattice structure undeformed and unrotated, ensuring the uniqueness of the decomposition in \eqref{eq:1}. The unique feature of CP theory is its construction of the plastic component $\Pb$ by constraining dislocation kinematics. Plastic flow evolves along pre-selected slip directions via volume-preserving shears, leaving the crystal lattice undistorted and stress-free \cite{McHugh2004-kl}.

Since in applications involving large deformations the elastic component of the deformation is small  with respect to the inelastic one, it can be neglected and   a rigid-visco-plastic approach will be adopted (such a hypothesis is generally used (\cite{lebtome93,kok02}). That means that  we  neglect  the elastic lattice strain $\Ub^e$ by supposing that  $\Ub^e \approx  \Ib $.  This leads to the following decomposition for the deformation gradient $\Fb$ (see for example \cite{kok02}): 
\begin{equation}
	\Fb=\Rb\Pb.
	\label{RP}
\end{equation}
Such a hypothesis is valid, as during forming or other industrial processes, the elastic component of deformation is negligibly small (typically on the order of 10$^{-3}$) compared to the plastic component (typically greater than 10$^{-1}$). It should also be noted that, once the elastic–plastic transition is complete, the stress evolution within the grains is governed by plastic relaxation (see  \cite{lebtome93}).

\subsubsection{Eulerian description  of  the lattice rotations}

Crystal slip systems are labeled by integers $s= 1,...,N$, with $N$ denoting  the number of slip systems. Each slip system $s$ is  specified by the unit vectors $(\bbs_s^0, \bm_s ^0)$, where $\bbs_s^0$ is in the slip direction and $\bm_s^0$ is normal to the slip plane in the perfect undeformed lattice.  Since the visco-plastic deformation does not produce distortion or rotation of the lattice, the mean lattice orientation is the same in the reference and intermediate configurations.  % and is specified by  $(\bbs_s^0, \bm_s^0 ), s= 1...N$.
Let   $\bbs_s=\bbs_s(t)$ and $\bm_s=\bm_s(t)$, be the glide direction and glide plane normal, respectively in the deformed configuration with  $\bbs_s(0)=\bbs_s^0$  and $\bm_s(0)=\bm_s^0$ . Since elastic effects are neglected,
%\begin{equation}\label{evlnoi}
$	\bbs_s = \Rb \bbs_s^0, \quad   \bm_s = \Rb \bm_s^0$ and we have 
%\end{equation}
%  
%Note that according to (\ref{evlnoi}), $\bbs_s$ and   $\bm_s$ are unit vectors. Furthermore,
\begin{equation}\label{prod}
	\bbs_s\otimes \bm_s = \Rb\left(\bbs_s^0 \otimes \bm_s^0\right)\Rb^T. 
\end{equation}
We seek  to express the lattice evolution equations only in terms of vector and tensor fields associated with the current configuration. Let  $\bu=\bu(t,x)$, the Eulerian velocity field,  $\Lb$ the velocity gradient,  $\D$ the rate of deformation,  and $\W$ the spin tensor
\begin{equation}\label{DW}
	\Lb=\Lb(\bu)=\nabla \bu, \quad  \D=\D(\bu)=(\nabla \bu)^{symm}, \quad \W=\W(\bu) =(\nabla \bu)^{skew}.
\end{equation}
Since the visco-plastic deformation is due to slip only  the slip contribution to the visco-plastic deformation is  (\cite{ric71})
\begin{equation}\label{flowrulenoi}
	\dot{\Pb}\Pb^{-1} = \sum_{s=1}^N\dot{\gamma}^{s}\bbs_s^0\otimes\bm_s^0,
\end{equation}
where  $\dot{\gamma}^{s}=\dot{\gamma}^{s}(t)$ is the visco-plastic shear rate on the slip system $s$.    If we  denote by  
\begin{equation}\label{MR}
	\M_s= \left(\bbs_s\otimes\bm_s\right)^{symm}, \quad \Qb_s=\left(\bbs_s\otimes\bm_s\right)^{skew}, 
\end{equation}
then,   using $\Lb= \dot{\Fb}\Fb^{-1}= \dot{\Rb}\;\Rb^{T} + \Rb\dot{\Pb}\Pb^{-1}\;\Rb^{T}$,  eqs.  (\ref{prod})   and  (\ref{flowrulenoi}), the rate of deformation $\D$   can be  written as
\begin{equation}\label{D}
	\D = \sum_{s=1}^N \dot{\gamma}^{s}\M_s. 
\end{equation}
Taking the  anti-symmetric  part of  $\Lb$,  we  obtain that the spin tensor is $\W=\dot{\Rb}\;\Rb^{T}+\sum_{s=1} ^N\dot{\gamma}^{s} \Qb_s$  and a differential equation for the rotation tensor $\Rb$:
\begin{equation}\label{Rp}
	\dot{\Rb}=(\W-\sum_{s=1} ^N\dot{\gamma}^{s} \Qb_s)\Rb. 
\end{equation} 
The evolution equations (\ref{Rp})  describe the evolution of the lattice in terms of vector and tensor fields associated with the current configuration.

\subsubsection{Plastic and  visco-plastic flow rules }
\label{flowruled}

In order to complete the model, we need to provide the constitutive equation for the slip rate $\dot{\gamma_s}$ as a function of $\tau_s$, the stress  component acting on the slip plane of normal 
$\bm_s$ in the slip direction $\bbs_s$.
In the current configuration, $\tau_s$ is expressed as
%\begin{equation}
$	\tau^s = \bs : \M_s$, 
%\end{equation}
where $\bs=\bs(t)$ is the Cauchy stress tensor acting in the current configuration $\K$ while $\M_s$ is defined by (\ref{MR}). Note that  $\{\tau^s\}_{s=\overline{1,N}}$ are not independent; they belong to a fifth dimensional subspace  of $\R^N$ corresponding to the dimension of the space  of deviatoric stresses. 

To determine the slip rates $\dot{\gamma}_{\alpha}$ relative to the local stress, a constitutive law is needed. Various proposals exist, ranging from phenomenological to more physically based approaches. One simple phenomenological approach assumes that $\dot{\gamma}_{\alpha}$ depends on the stress only through the resolved shear stress $\tau^s$.

In rigid-plastic formulations, it is assumed that the onset of plastic flow of a slip system $s$ is governed by Schmid law:  the slip system $s$ is active if and only if $\vert \tau^s \vert=\tau_c^s $, i.e., 
\begin{equation}\label{Schmid}
	\dot{\gamma_s}(\vert \tau^s \vert -\tau_c^s)=0, \quad \dot{\gamma_s} \tau^s \geq 0, \quad \vert \tau^s \vert -\tau_c^s \leq 0, 
\end{equation}
where  $\tau_c^s$ is the slip resistance (also called critical resolved shear stress or CRSS).  For a given time $t$, the $\tau_c^s$ are material constants. Thus, the planes $\vert \tau^s \vert=\tau_c^s $ are the facets of the current yield surface of the single crystal in the stress space.  

Since the resolved shear stresses $\tau^s$ are not independent, the shear rates $\dot{\gamma^s}$ given by the visco-plastic flow rule (\ref{Schmid}) are not independent; they have to satisfy the kinematic constraint (\ref{D}).  Given the rate of deformation $\D$, the slip rates $\dot{\gamma^s}$  can be determined  by minimizing the internal plastic dissipation power
\begin{equation}\label{workSchmid}
	J_{Schmid}(\dot{\gamma}^{1}, \dot{\gamma}^{2},...,\dot{\gamma}^{N})= \sum_{s=1}^{N}   \tau_c^s |\dot{\gamma}^{s}|,
\end{equation}
over $\R^N$ under the constraint (\ref{D}). The above functional is neither strongly convex  or  differentiable  and the solution could not be  unique.  %Additional assumptions are needed in order to restrict the number of solutions. 
One way to overcome the difficulty of determining the active slip systems and to take into account the viscous effects is to  adopt a rate-dependent approach for the constitutive response of the single crystal.
A widely used  rate-dependent (visco-plastic) model is the Norton type model, which  relates the shear strain rate $\dot{\gamma}^{s}$ on a slip system $s$  to the resolved shear stress   $\tau^s $  through a power-law   (see Asaro and Needleman \cite{asaneed85}) 
\begin{equation}\label{shear_strain1}
 	\dot{\gamma}^{s}=\dot{\gamma}_{0}^{s}\,{\left\vert\frac{\tau^s}{\tau_c^s}\right\vert}^n\,\mbox{sign}(\tau^s),
\end{equation}
where $\dot{\gamma}_{0}$ is a reference shear strain rate,  while the exponent $\emph{n}$ has a fixed value.    Since the internal plastic dissipation power reads
\begin{equation}\label{workNorton}
	J_{Norton}(\dot{\gamma}^{1}, \dot{\gamma}^{2},...,\dot{\gamma}^{N})=\frac{n}{n+1}  \sum_{s=1}^{N}  \tau_c^s |\dot{\gamma}^{s}| \left\vert\frac{ \dot{\gamma}^{s}}{\dot{\gamma}_{0}^{s}}\right\vert^{\frac{1}{n} },
\end{equation}
we remark that  $J_{Norton} \to 	J_{Schmid} $ for large values of $n$ ($n \to \infty$) which  means that the Norton law is a visoplastic regularization of  the Schmidt law.  Note that the plastic dissipation functional is strongly convex and differentiable meaning  the slip rates are the unique solution of the minimization problem of the  plastic dissipation power 	$J_{Norton}$  over $\R^N$ under the constraint (\ref{D}).

%\begin{figure}[ht]
%	\begin{center}
%		\includegraphics[scale=0.2]{Chap1/Images/FlowRuleMM_perzyna.pdf}
%	\end{center}
%	\caption{Schematic representation of  two visco-plastic regularizations of the %Schmid law (red line) : a power law (blue line)  and  overstress Perzyna-type %visco-plastic  law (black line).}
%	\label{Flowrule}
%\end{figure}

Another regularization  of the Schmid law  can be done by using a  Perzyna-like visco-plastic law  of the form. % (see also Figure \ref{Flowrule}), 
\begin{equation}\label{flow_rule}
	\dot{\gamma}^s = \dfrac{1}{\eta_s}\left[\vert \tau^s \vert-\tau_c^s \right]_+ \mbox{sign}(\tau^s), 
\end{equation}
where $\eta_s$ is the  viscosity, which may depend on the slip rate,  and $[ \;  x\; ]_+=(x+|x|)/2\;$ denotes the positive part of any real number $x$. 
Note that the visco-plastic flow rule (\ref{flow_rule}) is the visco-plastic extension of the rigid-plastic Schmid law using an over-stress approach.  The physical motivation for the dependence of the visco-plastic shear rate on the over-stress $(\tau^s-\tau_c^s)$ was provided by Teodosiu and Sidoroff \cite{TeodSid76} based on an analysis of the microdynamics of crystals defects. 

The internal power is given by 
\begin{equation}\label{workPerzyna}
	J_{Perzyna}(\dot{\gamma}^{1}, \dot{\gamma}^{2},...,\dot{\gamma}^{N})= \sum_{s=1}^{N}  \frac{\eta_s}{2}|\dot{\gamma}^{s}|^2 +  \tau_c^s |\dot{\gamma}^{s}|,
\end{equation}
and we remark that $	J_{Perzyna} \to 	J_{Schmid}$  for a vanishing viscosity $\eta_s$ ($\eta_s \to 0$).  The above functional is  strongly convex, hence the slip rates are the unique solution of  the minimization problem of the  plastic dissipation power   $J_{Perzyna}$ over $\R^N$ under the constraint (\ref{D}).

Finally we deduce that the   slip rates $\dot{\gamma}^{1}, \dot{\gamma}^{2},...,\dot{\gamma}^{N}$ can be obtained by an optimization problem involving the plastic dissipation power  
\begin{equation}\label{argMin}
	(\dot{\gamma}^{1}, \dot{\gamma}^{2},...,\dot{\gamma}^{N})= \arg \min_{\D = \sum_{s=1}^N g^{s}\M_s}	J(g^{1}, g^{2},...,g^{N}).,
\end{equation}
where  $J$ is one of the plastic dissipation power functionals    	$J_{Schmid}, 	J_{Norton}$ or   $J_{Perzyna}$.

%%%%%%%%%%%%%%%%%%%%%%%%

\subsection {Stability analysis  and attractors  of crystal orientation}

We will suppose in this section that we deal with a velocity gradient  driven processes, i.e., $\Lb=\Lb^*$ is  given and let us denote by  $\D^*=(\Lb^*+\Lb^{*T})/2$ the driven rate of deformation and by $\W^*=(\Lb^*-\Lb^{*T})/2$ the driven rotation tensor.

\subsubsection {Lattice orientation's Cauchy problem}

Bearing in mind that $\M_s=\Rb\M_s^0\Rb^T$  then the slip rate decomposition of  $\D^*$  (\ref{D})  can be  rewritten as  $\Rb^T\D^*\Rb=\sum_{s=1}^N \dot{\gamma}^{s}\M_s^0$. Hence, for a given  strain rate $\D^*$, the plastic (or visco-plastic) constitutive  law (\ref{argMin})    will give the slip rates $\dot{\gamma}^{s}$ as function of $\Rb$ only, i.e., 
\begin{equation}\label{argMindriven}
	(\dot{\gamma}^{1}(\Rb), \dot{\gamma}^{2}(\Rb),...,\dot{\gamma}^{N}(\Rb))= \arg \min_{\Rb^T\D^*\Rb = \sum_{s=1}^N g^{s}\M_s^0}	J(g^{1}, g^{2},...,g^{N}).
\end{equation}
We deduce that the differential equation(\ref{Rp})   is now a Cauchy problem for the rotation tensor $\Rb(t)$:  
\begin{equation}\label{PCRp}
	\left\{
	\begin{aligned}  &   	\dot{\Rb}(t)=\W^*\Rb(t)-\Rb(t)\left(\sum_{s=1} ^N\dot{\gamma}^{s}(\Rb(t)) \Qb_s^0\right), 
		\\ 
		& \Rb(0)=\Ib.
	\end{aligned}
	\right.
\end{equation} 

The main objective of this section is to give a  framework  for the stability analysis of the above differential  system and to  get some insights into possible attractors of the lattice rotation.   
%Let us first introduce the stationary lattice rotations $\Rb^{st}$ as the solution of the right hand side of the above evolution equation: 
%\begin{equation}\label{PCSt}
%	\W^*\Rb^{st}=\Rb^{st}\left(\sum_{s=1} ^N\dot{\gamma}^{s}(\Rb^{st}) \Qb_s^0\right). 
%\end{equation} 
%The above equation for the  stationary rotations is nonlinear and very difficult to solve in the set of  orthogonal tensors ${\cal O}=\{\Qb\; ;  \; \Qb^T=\Qb^{-1}\}$. 
The fact that  the rotation group ${\cal O}=\{\Qb\; ;  \; \Qb^T=\Qb^{-1}\}$ is not a vectorial space is  an important  difficulty which occurs when   we would like to use the linear stability theory to distinguish  between the stable and unstable stationary rotations.   That is why it could be more convenient  to  decompose    the  lattice rotation $\Rb$  in three elementary rotations  using  three  angles $\bth=(\theta_1,\theta_2,\theta_3)$, and denote it by $\Rb(\bth)$.    If we deal with   Tait–Bryan angles $(\theta_1,\theta_2,\theta_3)=(\alpha,\beta,\gamma)$ then  $\Rb(\bth)=\Rb(\theta_1-\theta_1^0,\be_1)\Rb(\theta_2-\theta_2^0,\be_2)\Rb(\theta_3-\theta_3^0,\be_3)$, where we have denoted by $\Rb(\theta_i,\be_i)$   the rotation with angle $\theta_i$ around the axis $Ox_i$ and by  $\bth^0=(\theta_1^0,\theta_2^0,\theta_3^0)$ the initial orientation of the crystal   with respect to  $Ox_1x_2x_3$  Eulerian coordinates. If we deal with  proper Euler angles $(\theta_1,\theta_2,\theta_3)=(\varphi,\theta,\psi)$  then $\Rb(\bth)=\Rb(\theta_1-\theta_1^0,\be_3)\Rb(\theta_2-\theta_2^0,\be_1)\Rb(\theta_3-\theta_3^0,\be_3)$. In all cases  $\dot{\gamma}^{s}$  are functions of $\bth$ through 
\begin{equation}\label{argMindriventheta}
	(\dot{\gamma}^{1}(\bth), \dot{\gamma}^{2}(\bth),...,\dot{\gamma}^{N}(\bth))= \arg \min_{\Rb(\bth)^T\D^*\Rb(\bth) = \sum_{s=1}^N g^{s}\M_s^0}	J(g^{1}, g^{2},...,g^{N}).
\end{equation}
If we denote by $\bOm=	\dot{\Rb}\Rb^T$ the  rotation rate tensor  then from  (\ref{PCRp})  we get: 
$$   	\bOm(t)=\W^*-\sum_{s=1} ^N\dot{\gamma}^{s}(\bth(t)) \Qb_s.$$ 
Since the above  equation involves only  skew tensors  it is more useful to   use the vector representation of a skew tensor. For  a skew tensor $\bA$ let $\ba$ be such that $\bA \br=\ba \wedge \br$ for all $\br$, or $a_1=A_{32}, a_2= A_{13},  a_3= A_{21}$. If we  denote by $\bom, \bw^*, \bq_s,  \bq_s^0$ the vectors associated to  $\bOm, \bW^*, \Qb_s, \Qb_s^0$ (note that $\bq_s=\bq_s(\bth)=\Rb(\bth)\bq_s^0$) then  the above equation reads 

\begin{equation}
	\label{Eqw}
	\bom(t)=\bw^*-\sum_{s=1} ^N\dot{\gamma}^{s}(\bth(t)) \bq_s(\bth(t)). 
\end{equation}
There exists a basis $\{\ba_1,\ba_2,\ba_3\}$, which could depend on $\bth$,  such that the angular velocity vector $\bom$ can be written as $\bom=\dot{\theta}_1\ba_1+\dot{\theta}_2\ba_2+\dot{\theta}_3\ba_3$. For instance, for the Euler angles  $\ba_1=\be_3, \ba_2=\be_N=\cos(\varphi)\be_1+\sin(\varphi)\be_2$ and $\ba_3=\sin(\varphi)\sin(\theta)\be_1-\cos(\varphi)\sin(\theta)\be_2+\cos(\theta)\be_3$.   Bearing in mind these notations equation (\ref{Eqw}) can be written now as a Cauchy problem in terms of   angles $\bth$ as 
\begin{equation}\label{PCth}
	\left\{
	\begin{aligned}  &   	\dot{\bth}(t)=\bFs(\bth(t)), 
		\\ 
		& \bth(0)=\bth^0,
	\end{aligned}
	\right.
\end{equation} 
where the three components $\F_i$  of  $\bFs$ are  given by:
$$\sum_{i=1} ^3\F_i(\bth)\ba_i(\bth)= \bw^*-\sum_{s=1} ^N\dot{\gamma}^{s}(\bth) \bq_s(\bth).$$

\subsubsection{Stability analysis of   velocity gradients driven processes} 

To describe the lattice orientation attractors, we first need to compute the stationary (or invariant) lattice orientation, denoted by $\bth^{st}$, as the solution to the following system:\begin{equation} \label{Satth}
	\bw^*=\sum_{s=1} ^N\dot{\gamma}^{s}(\bth^{st}) \bq_s(\bth^{st}). 
\end{equation}
Since the attractors are linearly stable stationary lattice orientations, the stability of $\bth^{st}$ can be characterized by the eigenvalues $\lambda_i(\bth^{st})$, with $i = 1, 2, 3$, of the $3 \times 3$ matrix:
$$ T_{ij}(\bth^{st}=\frac{\partial{\cal F}_i}{\partial\theta_j}(\bth^{st}).$$     
If 
\begin{equation} \label{Lamthb}	
	Re(\lambda_i(\bth^{st}))<0 \quad \mbox{for all} \; i=1,2,3  
\end{equation} 
then $\bth^{st}$ is linearly stable and is an attractor, denoted by $\bth^{att}$.  That means that there exists a  neighborhood ${\cal B}$  of  $\bth^{att}$, called  attraction basin, such that if  the initial orientation $\bth^0 \in {\cal B}$ then $\bth(t) \to  \bth^{att}$ for $t \to \infty$. 

\bigskip

Note that the time derivative in (\ref{PCth}) is material (also called particle or total derivative). This means that even if an Eulerian framework is used, the above stability analysis concerns the crystal orientation of a particle; in other words, it deals with a Lagrangian description.

\bigskip 

We must mention here that the application of the above stability analysis to a specific 3-D crystal (FCC, HCP, etc.) is quite complex. Indeed, it is very difficult to find the $N$ slip rate functions $\dot{\gamma}^s$ for the three angles $\bth$ from (\ref{argMindriventheta}) for a given rate of deformation $\D^*$. The second difficulty arises from solving the nonlinear system (\ref{Satth}). The 3-D case is beyond the scope of this paper, but it could be addressed in a  future work. In the next section, we will apply the same methodology to approach the 2-D case with three slip systems, where we were able to resolve the two main difficulties outlined earlier.

%Let us mention here that the  above  differential equation is generic, that means that  we are not able to get a   specific expression of the function  $\bFs$ in the general case. Only in some simple cases we are able to do it, as  for instance     (\ref{PCth2D}) obtained in the next  section. 

%%%%%%%%%%%%%%%%%%%%%%%%%

\section{Deeper analysis  of  2-D velocity gradient  driven processes}

The analysis of 3D crystalline systems with $N$ slip systems  presents significant theoretical and physical challenges. To facilitate understanding of the complex deformation phenomena, simplifications are often beneficial. 

\subsection{The 2-D model} \label{ToyModel}

\subsubsection{The in-plane  model with 3 slip systems}

  We will  recall  here   from \cite{Cazacu2010-hi}  {\color{blue} (see also \cite{lynggaard2019finite})} a two-dimensional  model   with $N=3$ slip systems.   Unlike the planar crystal model with 
$N=2$ slip systems proposed by Prantil et al. \cite{PJD93}, the 3-slip system model, used here, is not kinematically determined—meaning the distribution of shear across the slip systems is not predetermined.  
 Let us denote with $\phi$ and  $-\phi$   the angles  formed by slip system $r=1$ with the other two systems $r=2,3$ and let  $\theta$ be the  angle formed by the slip system $1$ with the $Ox_1$ axis, see Fig. \ref{G1G2G3P}a.   The three composite in-plane slip systems  $ \bbs_1, \bbs_2,  \bbs_3$ are  specified  by  the angles  $\theta,  \theta+\phi,  \theta-\phi$.
\label{sec:TOYMODEL}

%\begin{figure}[ht]
%	\begin{center} 
%		\includegraphics[scale=0.2]{Chap1/Images/3slip.pdf}
%	\end{center}
%	%\vspace{-1.5cm}
%	\caption{Two dimensional model with three slip systems.} \label{3slip}
%\end{figure}
 
 \bigskip  
 We have in mind two situations where the above model is physically relevant: (i) in-plane deformation of an FCC crystal, and (ii) slip in the basal plane of a hexagonal crystal. %We first examine the case of in-plane deformation in FCC crystals, and subsequently consider the second scenario.
 
 \textit{In-plane  deformation of  FCC crystals.} 
 Rice \cite{ric71} showed that certain pairs of the  three-dimensional systems that are potentially active need to combine in order to achieve plane-strain deformation.   For a FCC crystal, with 12 potentially active slip systems,    we consider  $Ox_3$ axis to be parallel to  $ [1 1 0] $ in the  crystal basis, which  means that   the plane-strain  plane $(Ox_1x_2)$ is   the plane  $[\bar{1}10]-[001]$.  Geometrical and mechanical constraints  (see details in  \cite{Cazacu2010-hi,lynggaard2019finite})  have to be satisfied such that  6 slip systems are inactive and the other 6 ones form  three pairs of   composite  systems  serving as  $N=3$  in-plane  systems.    %The in-plane system  $r=1$  is formed  by $(k,l)=(3,12)$,  $r=2$  is formed  by $(k,l)=(4,5)$,  and $r=3$  is formed  by $(k,l)=(7,8)$ (here $k,l$ are number of slipping directions for the 3-D FCC crystal).  We also suppose in what follows that the systems $1,2,6,9,10$ and $11$ are not active.  For all $r=1,2,3$ we have denoted by   $\bbs_r,  \bm_r$    the normalized projections of the   corresponding three-dimensional slip directions  and normal directions  $(k, l)$  onto the $x_1x_2$-plane. 
 The angle $\phi$  between  the slipping systems $1$ and $2$ is
 \begin{equation}\label{t123}
 	\phi=\arctan(\sqrt{2}) \approx 54.7^\circ.
 \end{equation}
 Since there are  some scalar factors between the first two components  of the in-plane systems and the 3-dimensional ones given by  
 %\begin{equation}\label{q123}
 $	q_1=\frac{1}{\sqrt{3}}, \quad q_2=q_3= \frac{\sqrt{3}}{2},$
 %\end{equation}
 the 2-D composite slipping rate $\dot{\gamma}^r$ corresponds to the  3-D  slipping rate $\dot{\gamma}^k$  multiplied by $2q_r$.  As it follows from \cite{Cazacu2010-hi}  the  2-D yield limit   $\tau_c^r$ corresponds to the  3-D yield limit   $\tau_c^k$  divided by $q_r$ while the  2-D   viscosity   $\eta^r$ corresponds to the  3-D viscosity  $\eta^k$  divided by $q_r^2$. 
 
 \textit{Slip in the basal plane of a hexagonal crystal.} 
 %\subsubsection{Slip in the basal plane of a hexagonal crystal}  
 Alternatively, the 2D model with three slip systems corresponds to hexagonal close-packed (HCP) crystals (such as Ti, Mg, Zr, etc.) under plane strain conditions, where the deformation plane is aligned with the basal plane (0001) of the hexagonal lattice. In this setting—experimentally studied in \cite{crepin1996}—plastic strain is primarily accommodated by the three prismatic slip systems, namely the $(10\bar{1}0)\langle1\bar{2}10\rangle$ slip family. Each of these slip systems can operate independently of the others, as their strain contributions are consistent with plane strain conditions. It is worth noting that combinations of other HCP slip systems from the basal or pyramidal families could, in principle, also produce plane strain. However, these alternatives would require significantly more energy and are thus not observed experimentally. For this reason, such slip families are not considered in the present work. The symmetry of the prismatic slip systems leads to 
 \begin{equation} 
 	\phi=\pi/3 = 60^\circ.
  \end{equation}

%\subsubsection{Lattice orientation equation} 
\bigskip

The main simplification  for the 2-D problem comes from the  latice rotation  $\Rb$ which is now a rotation  $\Rb(\theta, \be_3)$ with angle $\theta$ along $Ox_3$ axis and  we have 
\begin{equation}\label{Rb}
	\Qb_r=\dfrac{1}{2}\left((1,0)\otimes(0,1) -  (0,1)\otimes(1,0)\right).
\end{equation}
for all $r=1,2,3$ and  (\ref{Rp}) has  a much more simpler form 
\begin{equation}\label{theta}
	\dot{\theta} = \frac{\partial \theta}{\partial t} + \bu\cdot \nabla \theta= \frac{1}{2}\left( \sum_{r=1}^{3}\dot{\gamma}^{r} -(\frac{ \partial v_1}{ \partial x_2}-
	\frac{ \partial v_2}{ \partial x_1})\right),
\end{equation}
where $\bu$ is the  2-D Eulerian velocity field.

%\subsubsection{Specific applications} 

%\textcolor{blue}{
%%%%%%%%%%%%%%%%%%%%%%%%%%%%%%%%%%%%%%%%
\subsubsection{Slip rate decomposition of the strain rate} 
%%%%%%%%%%%%%%%%%%%%%%%%%%%%%%%%%%%%%%%ù
Another important simplification obtained in \cite{Cazacu2010-hi} for this model is the availability of analytical expressions for the slip rates on the individual composite systems, denoted by $\dot{\gamma}_{r}$, for a given rate of deformation $\D$. In the Eulerian basis $\be_1, \be_2$, this rate of deformation is expressed as
%\begin{equation} \label{Dcar}
$	\D=D_{11}(\be_1\otimes\be_1-\be_2\otimes\be_2)+D_{12}(\be_1\otimes\be_2+\be_2\otimes\be_1)$, 
%\end{equation}
 the  lattice orientation is given by $\theta$ and the 
slip rates $\dot{\gamma}_{r}$ ($r=1,2,3$)  can be determined by minimizing the (visco-)plastic power  $J$  under the constraints  (\ref{D}).   

As it follows from \cite{Cazacu2010-hi},  the  three  slip rates form a three dimensional vector which can be  decomposed in the basis 
 $\{\Sb^\phi, \Cb^\phi, \Zb^\phi\}$, given by $\Sb^\phi=(0,\sin(2\phi),-\sin(2\phi))$,  $\Cb^\phi=(1,\cos(2\phi), \cos(2\phi))
$  and $\Zb^\phi=\Sb^\phi  \wedge  \Cb^\phi=(-\sin(4\phi), \sin(2\phi), \sin(2\phi))$ as 
\begin{equation}\label{Gamth} 
	(\dot{\gamma}_1(\theta), \dot{\gamma}_2(\theta), \dot{\gamma}_3(\theta))=s(\theta)\Sb^\phi+c(\theta)\Cb^\phi+z(\theta)\Zb^\phi, 
\end{equation} 
where  $s$ and $c$ are  determined from  the kinematic constraints  (\ref{D}) to be 
\begin{equation} \label{proj}
	s(\theta)=-2\frac{\cos(2\theta)D_{11}+\sin(2\theta)D_{12}}{\vert \Sb^\phi \vert^2},  \quad \quad c(\theta)=2\frac{\cos(2\theta)D_{12}-\sin(2\theta)D_{11}}{\vert \Cb^\phi \vert^2},
\end{equation}
with   $\displaystyle \vert \Sb^\phi \vert^2= 2\sin^2(2\phi)$ and $\vert \Cb^\phi \vert^2=1+2\cos^2(2\phi)$.   Since the  slip rates  depends on   $z$ hence the internal power $J$  is a function of $z$ only, i.e., $J(\dot{\gamma}^{1}, \dot{\gamma}^{2}, \dot{\gamma}^{3})=: J(z)$. 

 For instance for the Schmid  model  the plastic dissipation power   $	z \to J_{Schmid}(z)$ is differentiable everywhere with the exception of three points, $z_1^\theta,z_2^\theta$ and $z_3^\theta$ where $s(\theta)S_r^\phi+c(\theta)C_r^\phi+zZ_r^
\phi =0$,  given by $$z_1^\theta=-\frac{c(\theta)}{\sin(4\phi)}, \quad z_2^\theta=\frac{s(\theta)\sin(2\phi)+c(\theta)\cos(2\phi)}{\sin(2\phi)}, \quad z_3^\theta=\frac{-s(\theta)\sin(2\phi)+c(\theta)\cos(2\phi)}{\sin(2\phi)}.$$
However, the left and right derivatives $J_{Schmid}'(z_r^\theta-), J_{Schmid}'(z_r^\theta+)$ exist  and can be computed  from 
$$J_{Schmid}'(z)=\sum_{r=1}^3 \tau_c^r \vert Z_r^\phi\vert \frac{z-z_r^\theta}{\vert z-z_r^\theta\vert}.$$
Since $z \to J_{Schmid}(z)$ is piece-wise linear the minimum of this  convex  function is   $z(\theta)=z_m^\theta $ where $z_m^\theta$ is such that  $J_{Schmid}(z_m^\theta )=\min_{r=1,2,3}J^{Schmid}(z_r^\theta )$ and we get 
\begin{equation}\label{zS}
	z(\theta) = z_r^\theta  \quad  \mbox{ if }  \quad J_{Schmid}'(z_r^\theta-)J_{Schmid}'(z_r^\theta+) \leq 0.
\end{equation}  
For the Perzyna visco-plastic regularization the expression of $J'(z)$  is:
$$J_{Perzyna}'(z)=A^\theta+B^\phi z+\sum_{r=1}^3 \tau_c^r \vert Z_r^\phi\vert \frac{z-z_r^\theta}{\vert z-z_r^\theta\vert}, $$ with $\displaystyle A^\theta=\sum_{r=1}^3\eta_r (s(\theta)S_r^\phi+c(\theta)C_r^\phi)Z_r^\phi, \; B^\phi=\sum_{r=1}^3\eta_r (Z_r^\phi)^2.$
We can now solve the equation  $J_{Perzyna}'(z)=0$ to get the analytical expression of $z(\theta)$
\begin{equation}\label{zP}
	z(\theta)= \left\{
	\begin{aligned}  &   z_r^\theta  \quad   \mbox{ if }  \quad J_{Perzyna}'(z_r^\theta-)J_{Perzyna}'(z_r^\theta+) \leq 0,
		\\ 
		& \frac{1}{B^\phi} \left[ -A^\theta + \sum_{r=1}^3 \tau_c^r \vert Z_r^\phi\vert \frac{J_{Perzyna}'(z_r^\theta+) }{\vert J_{Perzyna}'(z_r^\theta+) \vert} \right] \quad    \hbox{otherwise}.
	\end{aligned}
	\right.
\end{equation}

%\begin{figure}[ht]
%	\begin{center} 
%		\includegraphics[scale=0.2]{Chap1/Images/DPsi.pdf}
%	\end{center}
%	\vspace{-1.cm}
%	\caption{Angles of  the principal strain rate directions $\psi$ and  the lattice %orientations  $\theta$. } \label{DPsi}
%\end{figure}

\subsubsection{Reference slip  rates} 

\label{SlipRates}

Starting from the above decomposition of the strain rate into slip rates, we investigate here the dependence of the slip rates on the lattice orientation, which will be a key tool in the stability analysis developed in Section 4.

\begin{figure}[ht]
	\begin{center} 
	\includegraphics[scale=0.5]{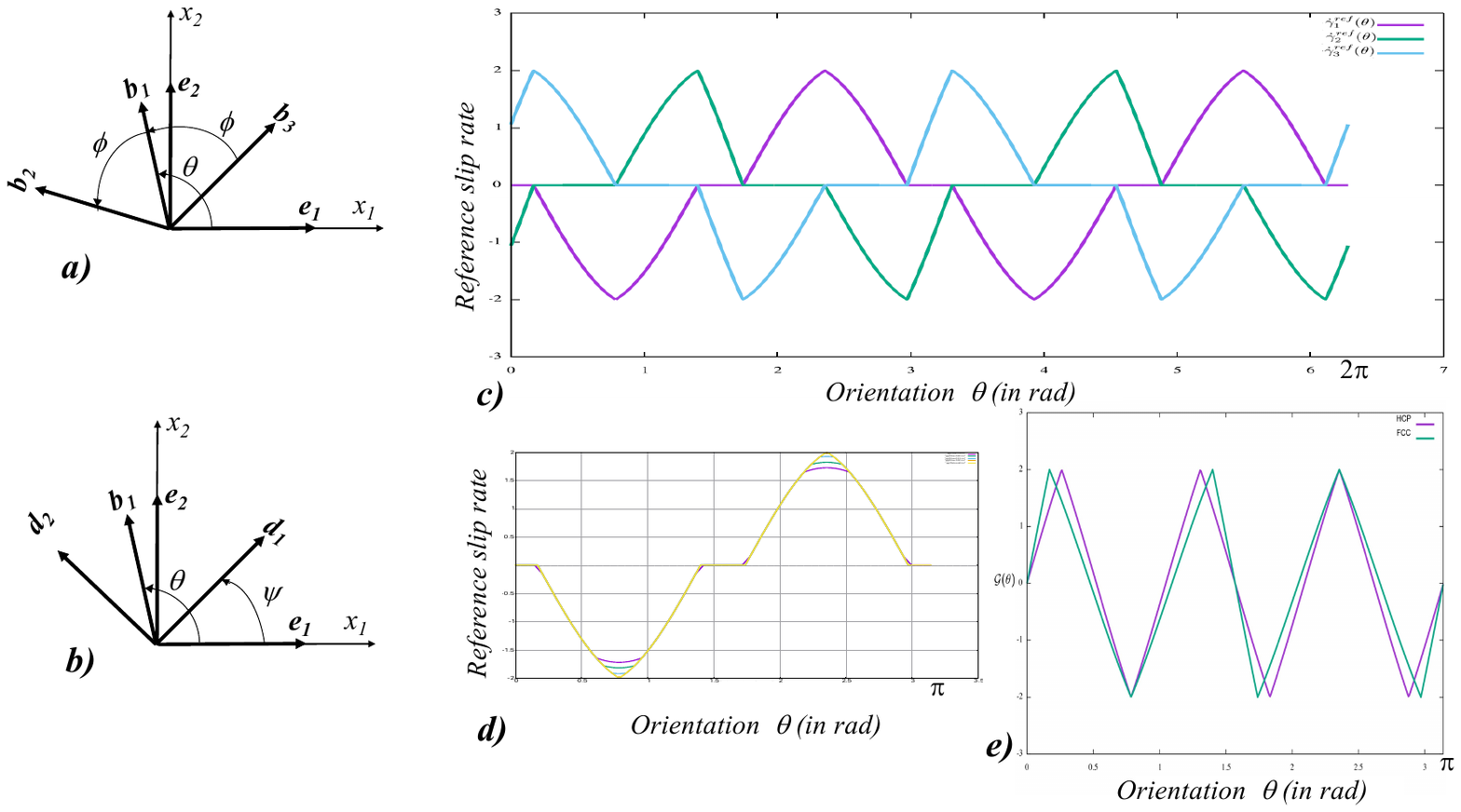}
	\end{center}
	    \vspace{-40pt} % adjust the value as needed (e.g., -5pt, -7pt)
	\caption{ (a) Two-dimensional model with three slip systems.
(b) Definition of angles: $\psi$, the principal strain rate direction, and $\theta$, the lattice orientation.
(c) Reference slip rates $\dot{\gamma}_1^{\text{ref}}(\theta)$, $\dot{\gamma}_2^{\text{ref}}(\theta)$, and $\dot{\gamma}_3^{\text{ref}}(\theta)$ as functions of $\theta$ for an FCC crystal (Schmidt model).
(d) Reference slip rate $\dot{\gamma}_1^{\text{ref}}(\theta)$ as a function of $\theta$ in the Perzyna model, for different values of the viscosity parameter: $\eta = 0.5\eta^{\text{ref}}$ (magenta), $\eta = 0.4\eta^{\text{ref}}$ (green), $\eta = 0.33\eta^{\text{ref}}$ (blue), $\eta = 0.30\eta^{\text{ref}}$ (red), and $\eta = 0$ (i.e., the Schmidt model, yellow), with $\eta^{\text{ref}} = \tau_c / \dot{\Gamma}^{\text{ref}}$.
(e) The function $\theta \mapsto \mathcal{G}(\theta)$ for a FCC crystal (green, $\phi = 54.7^\circ$) and a hexagonal crystal (blue, $\phi = 60^\circ$).} \label{G1G2G3P}
\end{figure}

The above formula, which gives the slip rate decomposition of the strain rate in the Eulerian basis,  is useful for the FE computations  and plays a key role in the numerical algorithm.  For the stability analysis is more convenient to represent the strain rate $\D$ in its principal directions,  denoted by $\bd_1,\bd_2$, as 
\begin{equation} \label{Dpd}
	\D=d(\bd_1\otimes\bd_1-\bd_2\otimes\bd_2),
\end{equation}
where $d>0$ is the principal rate of deformation. We denote  by  $\psi$ the angle between $\bd_1$ and $\be_1$ (see Fig. \ref{G1G2G3P}b).  To  compute the slip rate decomposition in this configuration let us consider first the non-dimensional case,  called in the following "reference case",  corresponding to  $D_{12}=0, D_{11}=1$ and to  $d=1$ and $\psi=0$.  Let denote by  $\dot{\gamma}_s^{ref}$   the "reference slip rates",   computed  with the formula of the previous subsection for $D_{12}=0, D_{11}=1$.  Let us mention here that  since  Schmidt  model is rate independent (i.e., the plastic dissipation power is a homogeneous functional of degree one),  the (dimensional) slip rate decomposition corresponding to the strain rate representation (\ref{Dpd})  can be recovered from  the reference slip rates as:
\begin{equation}\label{GamthRef} 
	(\dot{\gamma}_1(\theta), \dot{\gamma}_2(\theta), \dot{\gamma}_3(\theta))=d\left(\dot{\gamma}_1^{ref}(\theta-\psi), \dot{\gamma}_2^{ref}(\theta-\psi), \dot{\gamma}_3^{ref}(\theta-\psi)\right).
\end{equation}

In Fig. \ref{G1G2G3P}(c), we plot the reference slip rates $\dot{\gamma}_1^{ref}(\theta)$, $\dot{\gamma}_2^{ref}(\theta)$, and $\dot{\gamma}_3^{ref}(\theta)$ for the Schmidt model in the case of an FCC crystal. We note that for 12 orientations ($\theta = \phi/4, \pi/4, \pi/2-\phi/4, \pi/2+\phi/4, 3\pi/4, \pi-\phi/4, \dots$), only one slip system is active. For any other orientation, two slip systems are active, and the maximum/minimum of the reference slip rate is 2 and -2, respectively.

Moreover, the dependence of the slip rates on the lattice orientation is not smooth everywhere: at angles where only one slip system is active, the derivative is discontinuous. This derivative discontinuity, which is specific to the Schmid rigid-plastic law, is smoothed out in the case of Perzyna regularization. To illustrate this, we have plotted in Fig. \ref{G1G2G3P}d the first reference slip rate $\dot{\gamma}_1^{ref}$ as a function of the orientation $\theta$ for different values of the viscosity $\eta$, in the case of an FCC crystal. We observe that for larger viscosity values (greater than 50\% of $\tau_c/\dot{\Gamma}^{ref}$), the Schmid slip rate is not recovered by the Perzyna visco-plastic regularization, while for smaller viscosities (less than 25\% of $\tau_c/\dot{\Gamma}^{ref}$), the rigid-plastic and visco-plastic models become indistinguishable.

%\begin{figure}[ht]
%	\begin{center} 
%		\includegraphics[height=7.cm, angle=0]{Chap1/Images/gg1Vis.pdf}
%	\end{center}
%	\vspace{-0.25cm}
%	\caption{The reference slip rate $\theta  \to \dot{\gamma}_1^{ref}(\theta)$ as  functions of the lattice orientation  $\theta$  of  a  FCC crystal in-plane deformation for different  values of viscosity:  $\eta=0.5\eta^{ref}$ (in magenta),  $\eta=0.4\eta^{ref}$  (in green), $\eta=0.33\eta^{ref}$  (in blue),   $\eta=0.30\eta^{ref}$  (in red),  and  $\eta=0$ (the Schmidt model, in yellow) with  $\eta^{ref}=\tau_c/\dot{\Gamma}^{ref}$.} \label{Visco}
%\end{figure}

Moreover, it is convenient (see the next section) to define ${\cal G}$—the sum of the reference slip rates, as it appears in (\ref{theta})—given by:
\begin{equation} \label{G}
	{\cal G}(\theta)= 	\sum_{s=1} ^3\dot{\gamma}_{s}^{ref}(\theta),
\end{equation} 
In Fig. \ref{G1G2G3P}(e), we plot the non-dimensional function ${\cal G}$ for the in-plane deformation of an FCC crystal (green, $\phi = 54.7^\circ$) and of a hexagonal crystal (blue, $\phi = 60^\circ$).

\subsection {Stability analysis and attractors  } 

Applying the above stability analysis to determine the lattice orientation attractors is a challenging task. Indeed, deriving the expression for $\bFs$  in (\ref{PCth}) and solving the nonlinear system (\ref{Satth}) becomes very complex in the general case with numerous slip systems. However, for the 2-D model with three slip systems, many simplifications occur. Firstly, we deal with a single angle $\theta$ to describe the lattice orientation, and we have an analytical decomposition (\ref{Gamth}) instead of a minimization problem (\ref{argMindriven}).

\subsubsection {Lattice orientation's Cauchy problem} 

Let denote by $d^*>0$ and $-d^*$ the principal  rates of deformation, by  $\bd_1^*,\bd_2^*$  the    principal directions  (eigenvectors) of $\D^*$, by $\omega^*$ the  Eulerian rotation rate  and let    $\psi^*=\psi^*(t)$ be the angle  between  $\bd_1^*$  and  $Ox_1$ axis, i.e., we have 
$$\D^*=d^*(\bd_1^*\otimes\bd_1^* -\bd_2^*\otimes\bd_2^*),\quad \W^*=\omega^*(\be_1\otimes\be_2 -\be_2\otimes\be_1),  \quad \Lb^*=\D^*+\W^*. $$ 
%	where $d$ and $-d$ are the principal rates of deformations and $\omega $ is the rotation rate.  If the principal directions of the rate of deformation tensor $\D^*$ are given by the angle $\varphi^*$ then in what follows we have to replace $\theta$ by   $\theta-\varphi^*$. 

Let us deduce here ${\cal F}(\theta)$ the expressions of   $\bFs$  in (\ref{PCth}) for the  2D model.  For that let us remark  that from (\ref{GamthRef}) we get  
\begin{equation} \label{sum}
	\sum_{s=1} ^3\dot{\gamma}_{s} = d^*{\cal G}(\theta-\psi^*)
\end{equation}
where  ${\cal G}$  is  given  in (\ref{G})  and plotted in Figure \ref{G1G2G3P}e. Finally,  from (\ref{theta})  we deduce ${\cal F}(\theta)=\frac{1}{2}(d^*{\cal G}(\theta-\psi^*)-2\omega^*) $  and the 2D  expression of  (\ref{PCth}) :
\begin{equation}\label{PCth2D}
	\left\{
	\begin{aligned}  &   	\dot{\theta}(t)=\frac{1}{2}(d^*{\cal G}(\theta(t)-\psi^*(t))-2\omega^*) 
		\\ 
		& \theta(0)=\theta^0.
	\end{aligned}
	\right.
\end{equation}

%In the following, we consider two types of Eulerian strain-driven processes. The first involves a constant, i.e., stationary, velocity gradient with $\psi^(t) = \text{const}$. The second, referred to here as the "Lagrangian velocity gradient," has $\psi^(t) = \text{const} - \omega^*t$. In this case, the directions of the Eulerian strain gradient relative to the lattice remain constant over time.

\subsubsection{Stability analysis of  velocity gradients driven processes} 
\label{StabAnalStationary}

\begin{figure}[ht]
	\begin{center} 
		\includegraphics[scale=0.4]{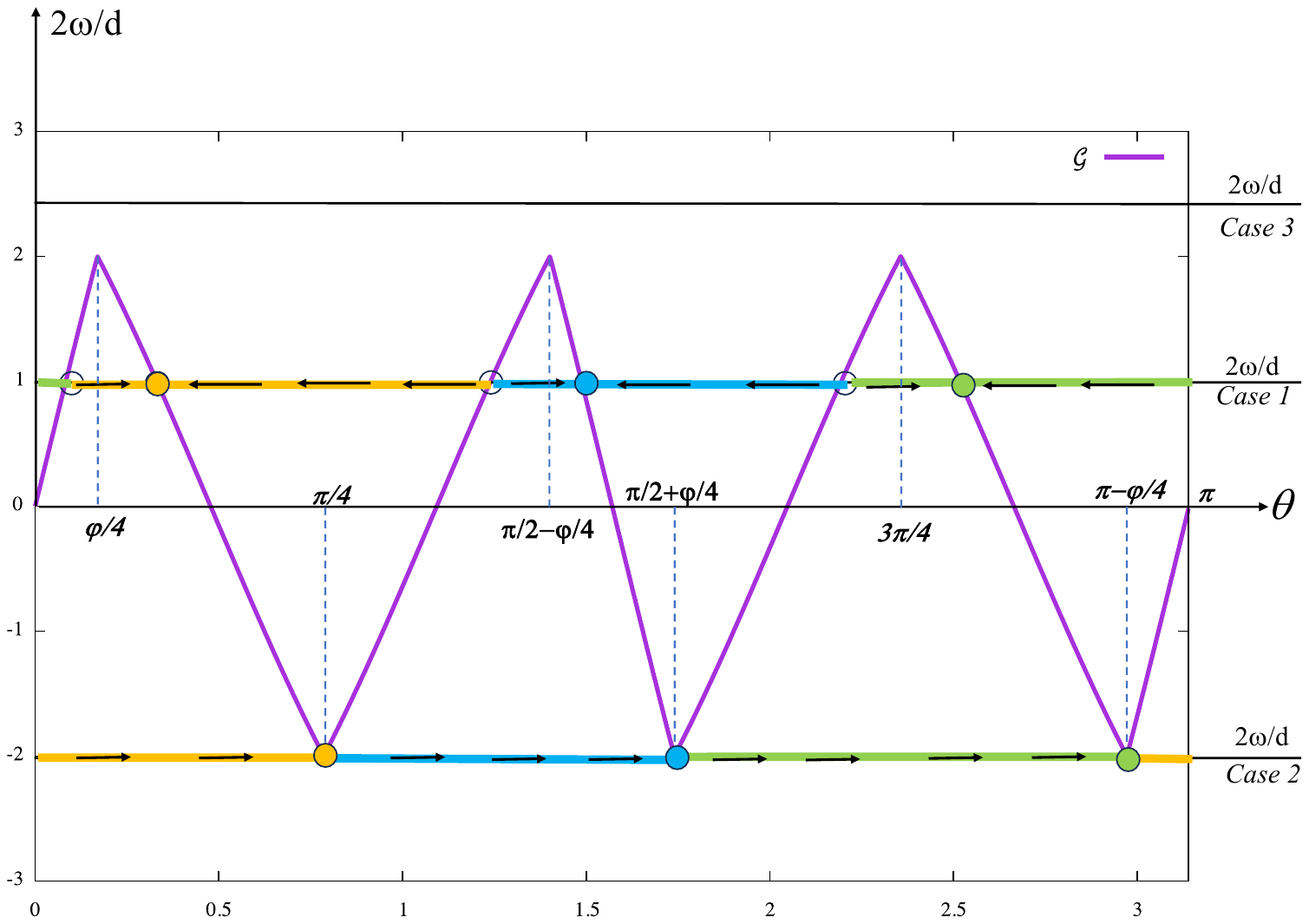}
	\end{center}
	\vspace{-30pt} % adjust the value as needed (e.g., -5pt, -7pt)	
	\caption{%\textcolor{blue}{
		The function $\theta \mapsto \mathcal{G}(\theta)$ in the case of in-plane deformation of an FCC crystal. Its intersection with the line $y = 2\tilde{\omega}^*/d^*$ defines the stationary lattice orientations $\tilde{\theta}^{\text{st}}$. The attractors $\tilde{\theta}^{\text{att}}$ are shown as solid colored discs (orange, blue, and green), and the colored segments (matching the color of each attractor) represent their respective basins of attraction.%}
	} \label{GP}
\end{figure}

For the sake of simplicity we will suppose in the next  that the angle $\psi^*(t)$  of  the principal directions of $\D^*$ has a constant rate  $ \dot{\psi}^*_0$, i.e.,  
$$
\psi^*(t)=	\psi^*_0+ \dot{\psi}^*_0 t,  
$$
and we denote by $ \tilde{\theta}(t)$  the  crystal orientation   with respect to  $\bd^*_1$ axis:  
$$
\tilde{\theta}(t)=\theta(t)-\psi^*(t). 
$$
Bearing in mind this notation,  (\ref{PCth2D}) reads
\begin{equation}\label{PCth2Dth}
	\left\{
	\begin{aligned}  &   	\dot{\tilde{\theta}}(t)=\frac{1}{2}({d^* \cal G}(\tilde{\theta}(t))-2\tilde{\omega}^*), 
		\\
		& 	\tilde{\theta}(0)=	\tilde{\theta}^0=\theta^0-\psi^*_0,
	\end{aligned}
	\right.
\end{equation} 
where $\tilde{\omega}^*={\omega}^*+\dot{\psi}^*_0$.  We can compute now the stationary orientations $\tilde{\theta}^{st}$ as the solution of the nonlinear algebraic equation 
\begin{equation} \label{eqT}
	{\cal G}(\tilde{\theta}^{st}) = 2\frac{\tilde{\omega}^*}{d^*}. 
\end{equation}
Let $\tilde{\theta}^{st}$ be a stationary orientation and suppose that ${\cal G}$ is differentiable in $\tilde{\theta}^{st}$. Then 
\begin{equation}\label{Stab}
	\mbox{if} \quad {\cal G}'(\tilde{\theta}^{st})<0  \quad  \mbox{then } \;   \tilde{\theta}^{st} \; \mbox{is an attractor.} 
\end{equation}	 
In Fig. \ref{GP} we have plotted the function $\theta \to {\cal G}(\theta)$   in the case of in-plane deformation of a  FCC crystal.  We remark that this function is piece-wise linear  and,  as it follows from (\ref{eqT}), the stationary   lattice orientation $\tilde{\theta}^{st}$  have to be founded at the intersection of this graph with the horizontal line $y=2\tilde{\omega}^*/d^*$. 

In what follows  we have chosen the time scale to be $1/d^*$ such that  the non-dimensional reference strain  $\gamma =d^* t$  has the same values as the time $t$.   We can distinguish 3 cases :

%\begin{itemize}  
%	\item  

i) $ \displaystyle \vert \tilde{\omega}^*\vert <   d^* $.  In this case, plotted in Figure  \ref{GP}a,  there are 6 stationary orientations  $\tilde{\theta}^{st}$  given by 
$$\tilde{\theta}_1^{st}=\frac{\tilde{\omega}^*}{d^*} \frac{\phi}{4},  \quad \tilde{\theta}_2^{st}=\frac{\phi}{4} +(1-\frac{\tilde{\omega}^*}{d^*}) \frac{\pi-\phi}{8}, \quad  \tilde{\theta}_3^{st}=\frac{\pi}{4} +(1+\frac{\tilde{\omega}^*}{d^*}) \frac{\pi-\phi}{8}, $$
$$\tilde{\theta}_4^{st}=\frac{\pi}{2} -\frac{\phi}{4} +(1-\frac{\tilde{\omega}^*}{d^*}) \frac{\phi}{4},  \quad \tilde{\theta}_5^{st}=\frac{\pi}{2} +\frac{\phi}{4} +(1+\frac{\tilde{\omega}^*}{d^*}) \frac{\pi-\phi}{8},\quad \tilde{\theta}_6^{st}=\frac{3\pi}{4} +(1-\frac{\tilde{\omega}^*}{d^*}) \frac{\pi-\phi}{8}$$
but following (\ref{Stab})  only three  ($\tilde{\theta}_2^{st}=\tilde{\theta}_1^{att}, \tilde{\theta}_4^{st}=\tilde{\theta}_2^{att}$ and $\tilde{\theta}_6^{st}=\tilde{\theta}_3^{att}$)   are attractors (stable):
\begin{equation}\label{Att}
	\tilde{\theta}_1^{att}=\frac{\phi}{4} +(1-\frac{\tilde{\omega}^*}{d^*}) \frac{\pi-\phi}{8},   \; 	\tilde{\theta}_2^{att}=\frac{\pi}{2} -\frac{\phi}{4} +(1-\frac{\tilde{\omega}^*}{d^*}) \frac{\phi}{4},  \;   \tilde{\theta}_3^{att}=\frac{3\pi}{4} +(1-\frac{\tilde{\omega}^*}{d^*}) \frac{\pi-\phi}{8},
\end{equation}
which are plotted in Figure \ref{GP} by solid  colored (orange, blue and green) discs.   
%\end{itemize}  
Their attraction basins are the intervals plotted in Fig. \ref{GP}  by colored segments (corresponding to the color of the attractor) given by  
$$ B^{att}_1=(\tilde{\theta}_1^{st}, \tilde{\theta}_3^{st}), \quad B^{att}_2=(\tilde{\theta}_3^{st},  \tilde{\theta}_5^{st}),  \quad B^{att}_3=(\tilde{\theta}_5^{st}, \pi+\tilde{\theta}_1^{st}).$$ 
This means that 
\begin{equation}\label{Attractor}
	\mbox{if} \quad \tilde{\theta}_0 \in  B^{att}_i \quad \mbox{then } \quad  \tilde{\theta}(t) \to \tilde{\theta}_i^{att}  \quad  \mbox{for }  \quad  t \to \infty, \quad   \mbox{for  all}  \; i=1,2,3.
\end{equation}

\begin{figure}[ht]
	\begin{center} 
		\includegraphics[height=4.5cm, angle=0]{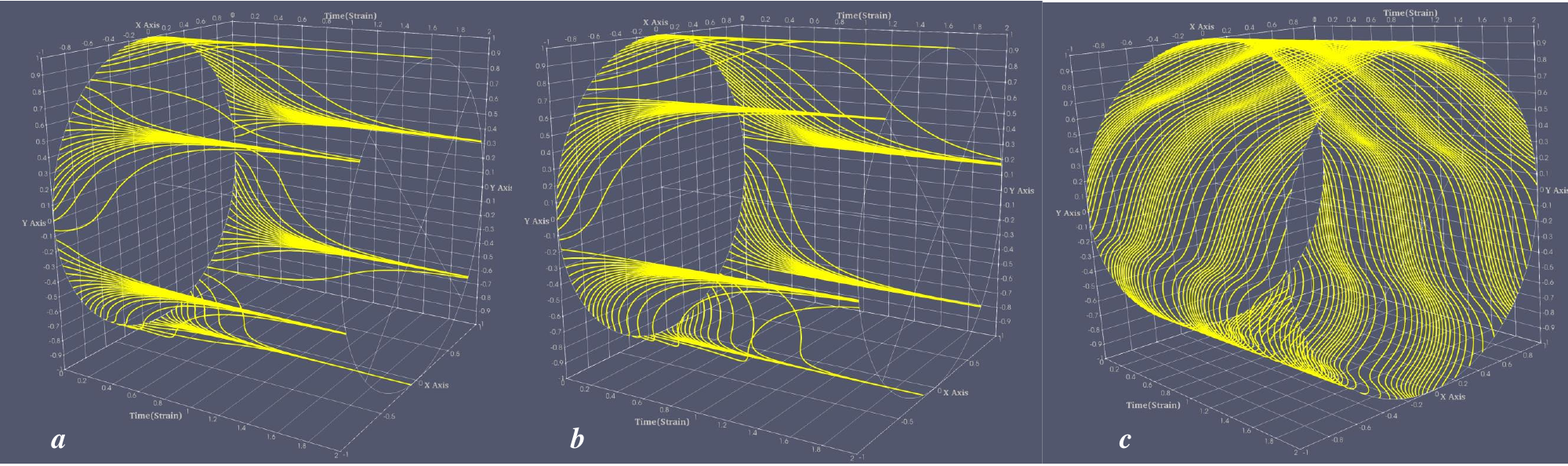}
	\end{center}
	\vspace{-.5cm}
	\caption{Numerical simulation  of  200  time/strain trajectories $t \to (t,\cos(\theta(t)), \sin(\theta(t)))$  of the crystal orientation for  different  choices of the initial orientation $\theta_0$ and $\psi^*=0$. Left: case i) ($ \displaystyle \vert \omega^*\vert <  d^*$) with  3  attractors.  Middle: case ii)  ($  -\omega^* =   d^*$) with 3 half-attractors. Right: case iii)  ($ \displaystyle \vert \omega^*\vert >  d^* $) with no attractors.} \label{OE1}
\end{figure}

In Fig. \ref{OE1}a, we plot the numerical simulation of the time evolution of the lattice orientations, $t \to (\cos(\theta(t)), \sin(\theta(t)))$, for 200 different initial orientations $\theta_0 \in [0,2\pi]$ and $\psi^*_0=0$. We have chosen $\omega^*= d^*/2$, corresponding to case 1 of Fig. \ref{GP}. We observe the presence of 3 different attractors (6 in Fig. \ref{OE1}a, with a  $\pi$ symmetry), as described by the theoretical results above. Note that the attractors are only valid for large strains, i.e., the distance $\vert \theta(t) - \theta_i^{att}\vert$ is small enough only for strains (slips $\gamma$) larger than $50\%$. These large strains occur frequently, even for small overall strains.

\bigskip

ii) $ \displaystyle \vert \tilde{\omega}^*\vert =  d^* $.  In this case, plotted in Fig.  \ref{GP},   which is specific to the slip bands (see section \ref{SlipBandsSection}),  there are 3 stationary orientations  $\tilde{\theta}$  given by 
$$\tilde{\theta}_1^{st}= \mbox{sign}(\frac{\tilde{\omega}^*}{d^*}) \frac{\phi}{4},  \quad \tilde{\theta}_2^{st}=\frac{\pi}{2}+ \mbox{sign}(\frac{\tilde{\omega}^*}{d^*}) \frac{\pi}{4} , \quad  \tilde{\theta}_3^{st}=\frac{\pi}{2}-  \mbox{sign}(\frac{\tilde{\omega}^*}{d^*})\frac{\phi}{4}.$$
For all these angles the function $\theta \to \  {\cal G}(\theta)$ is not differentiable  but   the left and right derivative exists and  have opposite signs. Following (\ref{Stab})   neither are stable (in the classical sense) but they have left or right attractors. That is why we will call them  "half-attractors". Indeed, for $\tilde{\omega}^*=d^*$    we have 
\begin{equation}\label{Att2p}
	\tilde{\theta}_3^{att}=\frac{\phi}{4},\quad 	\tilde{\theta}_2^{att}=\frac{\pi}{2}- \frac{\phi}{4},   \quad  \tilde{\theta}_1^{att}=\frac{3\pi}{4},  \quad \mbox{for} \; \tilde{\omega}^*=d^*, 	
\end{equation} 
with the attraction basins 
\begin{equation}\label{Att2pB}
	B^{att}_3=[\frac{\phi}{4},  \frac{\pi}{2}- \frac{\phi}{4}), \quad  B^{att}_2=[\frac{\pi}{2}- \frac{\phi}{4},\frac{3\pi}{4}), \quad     B^{att}_1=[\frac{3\pi}{4}, \pi+ \frac{\phi}{4}),   \quad \mbox{for} \; \tilde{\omega}^*=d^*. 
\end{equation} 
It should be noted that in the case $ \displaystyle \vert \tilde{\omega}^*\vert = d^* $, the attractor orientations correspond to specific configurations where only a single slip system is active (see section \ref{SlipRates}). The subscript $r$ of each attractor $\tilde{\theta}_r^{att}$ indicates which slip system is active: specifically, $\dot{\gamma}^r\neq 0$ while $\dot{\gamma}^s=0$ for all $s\neq r$.
\bigskip 

For  $d^*=-\tilde{\omega}^*$ the half-attractors  
\begin{equation}\label{Att2m}
	\tilde{\theta}_1^{att}=\frac{\pi}{4},\quad 	\tilde{\theta}_3^{att}=\frac{\pi}{2}+ \frac{\phi}{4},  \quad   \tilde{\theta}_2^{att}=\pi - \frac{\phi}{4},  \quad \mbox{for} \; \tilde{\omega}^*=-d^*, 	
\end{equation}
are plotted in Fig. \ref{GP} by solid  colored (orange, blue and green) discs,  while their attraction basins  
\begin{equation}\label{Att2mB}
	B^{att}_1=(-\frac{\phi}{4}, \frac{\pi}{4}], \quad B^{att}_3=(\frac{\pi}{4}, \frac{\pi}{2}+ \frac{\phi}{4}],  \quad B^{att}_2=(\frac{\pi}{2}+ \frac{\phi}{4}, \pi - \frac{\phi}{4}],  \quad \mbox{for} \; \tilde{\omega}^*=-d^*,
\end{equation}
are the intervals plotted  by colored segments (corresponding to the color of the attractor).

The  time evolution of the lattice orientations, $t \to (\cos(\theta(t)), \sin(\theta(t)))$, for 200 different initial orientations $\theta_0 \in [0,2\pi]$ and $\psi^*=0$  for $d^*=-\omega^*$ (corresponding to  case 2 of Fig. \ref{GP}) is plotted in Fig. \ref{OE1}b.   We remark the presence of 3  attractors and the  lattice  orientations  trajectories  obey   (\ref{Attractor}).  As before the attractors work only for large strains.

%We have to mention here that  in the case  $ \displaystyle \vert \omega^*\vert =  d^* $ the attractors orientation correspond to the specific orientations where {\em only one system is active} (see  section \ref{SlipRates}).  %The label $r$ of each attractor  $\tilde{\theta}_r^{att}$ was chosen such that {\em only the system $r$ is active} (i.e. $\dot{\gamma}^r\neq 0$ and $\dot{\gamma}^s=0$ for $s\neq r$). 

%\begin{figure}[ht]
%	\begin{center} 
%		\includegraphics[height=12.5cm, angle=0]{Chap2/Images/OEo=1.jpg}
%	\end{center}
%	\vspace{-.5cm}
%	\caption{Numerical simulation  of  200  time/strain trajectories $t \to (\cos(\theta(t)), \sin(\theta(t)))$  of the crystal orientation  in  case 2  ($  -\omega^* =   d^*$ and $\psi_0^*=0$) for  different  choices of the initial orientation $\theta_0$.  For  crystal symmetry reasons  only  3  (left) attractors  of  6  plotted here are different.  } \label{OE2}
%\end{figure}

\bigskip

iii) $ \displaystyle \vert \tilde{\omega}^*\vert >  d $.  In this case, plotted in Fig.  \ref{GP},  there are no  stationary orientations. To see that  we have chosen   $\displaystyle \tilde{\omega}^*=\frac{3}{2} d^*=1.5$ (corresponding to  case 3 of Fig. \ref{GP}) and we have plotted  in Fig. \ref{OE1}c the numerical simulation of the   time evolution of the lattice  orientations  $t \to (\cos(\theta(t)), \sin(\theta(t)))$  for 200 choices of the initial orientation $\theta_0 \in [0,2\pi] $  and  $\psi^*=0$.    We remark  the absence of any attractor.

%\begin{figure}[ht] 
%	\begin{center}
%		\includegraphics[height=12.cm, angle=0]{Chap2/Images/OEo=15.jpg}
%	\end{center}
%	\vspace{-.5cm}
%	\caption{Numerical simulation  of  200  time/strain trajectories $t \to (\cos(\theta(t)), \sin(\theta(t)))$  of the crystal orientation  in  case 3  ($ \displaystyle \vert \omega\vert >  d $ and $\psi^*_0=0$) for  different  choices of the initial orientation $\theta_0$.  Note that no attractors are present. } \label{OE3}
%\end{figure}

\section{Prediction of ideal orientations in classical problems }

 In this section we want to illustrate the above stability analysis  of the crystal lattice orientation evolution with   three different  classical problems of crystal plasticity.  The first one   deals  with a poly-crystal (i.e., a non-homogeneous initial orientation) which is loaded with an homogeneous  gradient velocity loading.  The second one concerns a mono-crystal which is subjected to an isotropic loading, as a void growth  under radial expansion. In this case the  associated rate of deformation $\D^*$ is radial, hence non-homogeneous in space while the initial orientation is homogeneous.   The third one deals with a slip band deformation in a uni-axial loading of a mono-crystal. 

 In all these cases, the theoretical results obtained for specific velocity-gradient-driven processes are compared to the numerical results of the associated boundary-value problems.  We have to mention here that   only  the boundary conditions  are given in relation to a driven  velocity field $\bu^*$  but the  computed velocity field $\bu$ is different from the expected one. Indeed the response of the (poly-)crystal  on the boundary loading,  could presents  slip and/or kink bands, which cannot be described by the  driven velocity field $\bu^*$. However, in the examples shown here,  {\em  the overall final orientation of the lattice can be  deduced from the apriori attractors distribution} predicted by  the stability analysis of the previous section.

\subsection{Settings and strategy}  

\subsubsection{Initial and boundary value problem formulation}

\label{BVP}

We begin by presenting the equations governing the  motion in a domain $\DD=\DD(t)$  of an incompressible rigid-visco-plastic crystal for the simpified 2D model with $N=3$ slip systems.  In an Eulerian description of a crystal visco-plasticity theory, the unknowns    are: the velocity $\bu: [0,T]\times \DD \to \R^2$, the crystal lattice orientation, i.e., the rotation $\theta: [0,T]\times \DD \to \R$  and the Cauchy stress $\bs : [0,T] \times \DD \to \R^{2\times2}_S$. Let   $\bs =\bs'+p\I$, where $\bs'$ is the stress deviator while $p : [0,T] \times \DD \to \R$ is the pressure (mean stress).

The momentum balance  in the Eulerian coordinates reads
\begin{equation} \label{mbl}
	\rho^{mass}(\partial_t \bu + \bu \cdot \nnabla \bu)  - \mbox{\bf div} \bs'  + \nabla p= \rho^{mass}{\ff} \quad
	\hbox{ in } \DD,
\end{equation}
where   the mass density $\rho^{mass} >0$ and the body forces $\ff$ are supposed to be known. The  incompressibility condition is
\begin{equation}\label{inc}
	\mbox{div}\;  \bu =0 \quad \hbox{ in } \DD.
\end{equation}
The momentum balance equations are completed by the  constitutive equation,  which relates the stress tensor $\bs$ and the rate of deformation tensor $\D(\bu)$ (see (\ref{D})) through   the evolution equations for each slip system $s$  given by (\ref{Schmid}) or some viscous regularization (\ref{shear_strain1}) or (\ref{flow_rule}).

The boundary $\partial \DD$ of the domain  $\DD$  is decomposed into two disjoint parts, $\Gamma_v$ and $\Gamma_s$, such that the velocity is prescribed on     $\Gamma_v$ and traction is prescribed   on  $\Gamma_s$, at any time $t$:
\begin{equation} \label{bcv}
	\bu(t) = {\boldsymbol V}(t) \quad \hbox{on} \quad \Gamma_v, \quad \bs(t) \bn = \Sb(t)
	\quad \hbox{on} \quad
	\Gamma_s,
\end{equation}
where $\bn$ stands for the outward unit normal on $\partial \DD$,
$\bV$  is the imposed velocity and $\Sb$ is the prescribed
stress vector.

We  also consider another partition of  $\partial \DD$
into $\partial_{in} \DD(t)$ and $\partial_{out} \DD(t)$
corresponding to  incoming ($\bu\cdot \bn <0$)  and outcoming  ($\bu\cdot \bn \geq 0$)  flux.  The boundary
conditions associated with  the lattice  evolution  equations  (\ref{theta}) reads 
\begin{equation}\label{bcbmTM}
	\theta(t) = \theta^{in}(t),   \quad     \quad \hbox{on} \quad
	\partial_{in} \DD(t),
\end{equation}
To  the field equations, we add the initial conditions
\begin{equation}\label{ic}
	\bu(0)=\bu^0, \quad \theta(0) = \theta^0,  \quad   \hbox{in} \quad
	\DD.
\end{equation}
where $\bu^0$ is the initial velocity and  $\theta^0$ gives the initial orientation of  the crystal lattice.  
%Note that in this model,  there is no need to prescribe initial conditions for the stress. This is very convenient since the initial stress field is generally not known or it cannot be easily measured.  

\subsubsection{Numerical strategy}
To solve the governing equations, we employ a hybrid computational approach that combines mixed finite element and discontinuous Galerkin methods, originally developed in \cite{CI09}. Below, we outline the key principles underlying our numerical simulations. Our approach uses an implicit (backward) Euler time discretization scheme for the field equations, leading to a system of nonlinear equations for the velocity field $\bu$ and the lattice orientation $\theta$. At each time step, we implement an iterative algorithm to solve these nonlinear equations using a two-part strategy: (i) the velocity field is treated using a variational formulation discretized via the finite element method, (ii) the lattice orientation is solved using a discontinuous Galerkin method with an upwind flux.

A significant advantage of this numerical framework is that it circumvents the need to model elastic deformation, as it does not rely on an elastic predictor/plastic corrector scheme, such as the one used in \cite{SH}. The rigid-visco-plastic model (\ref{flow_rule}) poses particular computational challenges due to the non-differentiability of its visco-plastic terms, rendering standard finite element techniques for Navier–Stokes fluids  unsuitable. We address this issue through a modified iterative decomposition–coordination formulation, coupled with the augmented Lagrangian method \cite{GlLT}. This modification is essential because, unlike the Bingham model  (von-Mises plasticity) studied in \cite{GlLT}, the crystal model (\ref{Schmid}) lacks co-axiality between the stress deviator and the rate of deformation.
%Our adapted algorithm effectively solves for the unit vectors defining lattice orientation at each iteration.}

For simulations where the domain $\DD$ evolves over time, we implement an Arbitrary Eulerian-Lagrangian (ALE) description. In this cas. To avoid the interpolation of the lattice orientation  on the deformed mesh,  is more convenient to have the same finite element Galerkin discontinuous meshes.   The numerical algorithm proposed here uses only  a Stokes-type problem at each time step and the  implementation of the Navier-Stokes equations in an ALE formulation is rather standard  (see for instance
\cite{Jafari2016}). Throughout the numerical studies presented in subsequent sections, we employ a consistent spatial discretization scheme: P2 elements for the velocity field, P1 elements for pressure, and P1 discontinuous elements for the stress field, slip rates, dislocation densities, and lattice orientations.

During large deformations, the aspect ratio of the elements can become excessively large, which may introduce errors and hinder strain localization. To mitigate this, a multi-step mesh refinement strategy is applied at each remeshing stage, adjusting the mesh density as deformation progresses. This gradual mesh refinement enhances the simulation resolution during deformation, minimizes information loss, and avoids a significant increase in computational cost, see \cite{resk2009adaptive}. In our adaptive meshing approach for both monocrystal and polycrystal simulations, we apply three combined criteria: strain gradient, orientation gradient, and accumulated plastic strain.

\subsubsection{Material settings}

The material parameters used in this quasi-static computation is not specific to a particular material but instead reflects the order of magnitude typically associated with hexagonal close-packed (HCP) crystals ($\phi = \pi/3 = 60^\circ$) for the first two simulations and with a face-centered cubic (FCC) crystal ($\phi = 54.7^\circ$) for the third. The material parameters are as follows: density $\rho^{\text{mass}} = 3200 \, \text{kg/m}^3$, yield limit $\tau_c^s = 20 \, \text{MPa}$ for HCP, and $\rho^{\text{mass}} = 16650 \, \text{kg/m}^3$, $\tau_c^s = 220 \, \text{MPa}$ for FCC. In all cases, no hardening effects were considered, and a small viscosity (less than $1\%$ of $\eta^{ref} = \tau_c^s / \dot{\Gamma}_c$, where $\dot{\Gamma}_c$ is the characteristic strain rate) was used for numerical reasons.

\subsection{Poly-crystal response to homogenenous loading}

\begin{figure}
	\center
%	\makebox[\textwidth][c]{\includegraphics[scale=0.4]{Chap2/Images/Polycrystal.pdf}}
\includegraphics[height=7cm, angle=0]{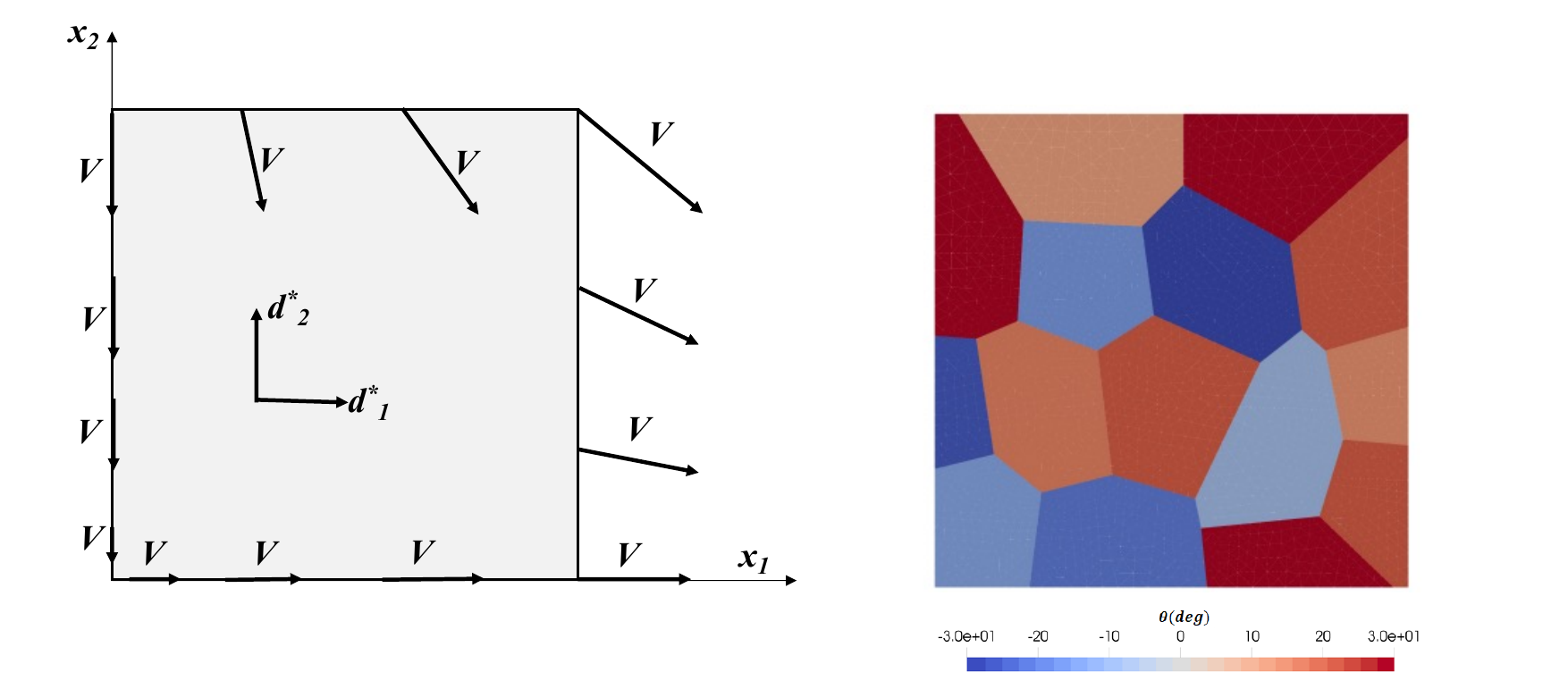}
	\caption{%\textcolor{blue}{
			Left: Schematic representation of a polycrystal subjected to a homogeneous velocity gradient loading.
Right: Spatial distribution $(x_1, x_2) \mapsto \theta^0(x_1, x_2)$ of the initial lattice orientation, shown in a color scale ranging from $-30^\circ$ to $30^\circ$.%}
}
	\label{fig:Poly}
\end{figure}

The  poly-crystal domain  is  a rectangle $(0,L(t))\times((0,H(t))$,  which is  a  square   ($L_0=H_0$) at the initial configuration $t=0$, containing 15 grains of different orientations (see Fig. \ref{fig:Poly}  right). 	
\begin{figure}
%	\center
%	\makebox[\textwidth][c]{\includegraphics[scale=0.8]
\includegraphics[height=10cm, angle=0]{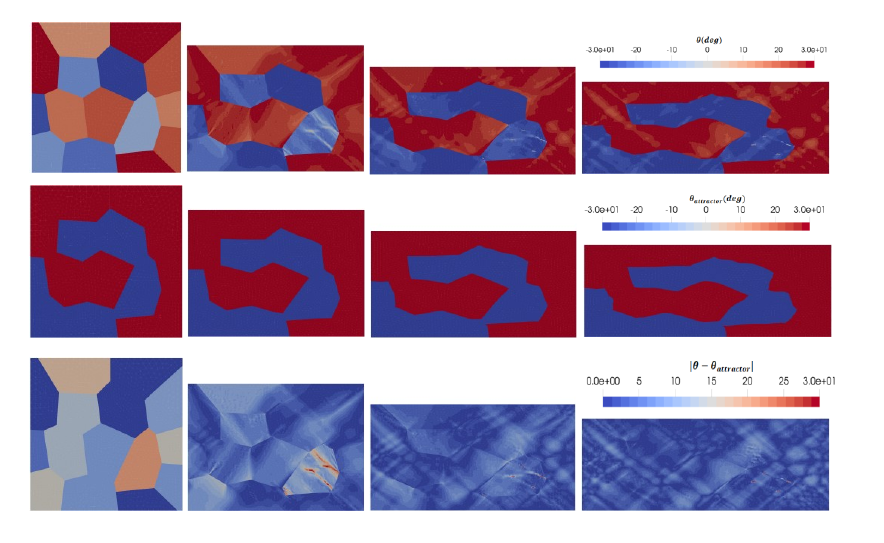}
	\caption{
		%\textcolor{blue}{
Top: Computed Eulerian distribution $(x_1, x_2) \mapsto \theta(t, x_1, x_2)$ of the lattice orientation, shown in a color scale ranging from $-30^\circ$ to $30^\circ$.
Middle: Expected Eulerian distribution $(x_1, x_2) \mapsto \theta^{\text{att}}(t, x_1, x_2)$ of the orientation attractor, in the same color scale $(-30^\circ, 30^\circ)$.
Bottom: Eulerian distribution $(x_1, x_2) \mapsto \vert \theta(t, x_1, x_2) - \theta^{\text{att}}(x_1, x_2) \vert$ representing the difference between the lattice orientation and the estimated attractor, shown in a color scale from $0^\circ$ to $30^\circ$.
The distributions are shown for four time points: $t = 0$ ($\epsilon^{\text{eng}} = 0$), $t = T/3$ ($\epsilon^{\text{eng}} = 0.166$), $t = 2T/3$ ($\epsilon^{\text{eng}} = 0.33$), and $t = T$ ($\epsilon^{\text{eng}} = 0.5$).%}
}
	\label{fig:PolyOrient}
\end{figure}

On the  boundary, denoted by $\Gamma_v(t)=\Gamma(t)$, we impose a   velocity $\bV$ (see Fig. \ref{fig:Poly}  left), corresponding to an  in-compressible Eulerian  velocity field $\displaystyle \bu^*=d^*(x_1\be_1-x_2\be_2)$, i.e., 
\begin{equation}
	{\bV}(t)(x_1,x_2)= d^*(x_1\be_1-x_2\be_2).  
\end{equation} 
The associated gradient of the velocity field $\bu^*$  is 
$\Lb^*=\D^*=d^*(\be_1\otimes\be_1 -\be_2\otimes\be_2 ),$ 
which is homogeneous and corresponds to the  choice 
$\bd_1^*=\be_1, \quad \bd_2^*=\be_2, \quad \omega^*=0, \quad \psi^*=\psi_0^*=0.$ The initial  velocity and  the  initial   crystal orientation where chosen to be 
\begin{equation}
	\bu^0 (x_1,x_2)= \bu^*(x_1,x_2), \quad \theta^0=\theta^0(x_1,x_2). 
\end{equation}

Following the stability analysis of the  rotated orientation angle $\tilde{\theta}=\theta$ we deduce that the orientation attractor (plotted in Fig. \ref{fig:PolyOrient} middle)  is given on each   grain  depending on the initial orientation  $ \theta^0$ (plotted in  Fig. \ref{fig:Poly} right) by 
$$\theta^{att}(x,y) =
\left\{
\begin{aligned}   
	\frac{\pi}{6} &\quad \mbox{if} \;  &  \theta^0(x,y)& \in (0,\frac{\pi}{3})	\\ 
	-\frac{\pi}{6}& \quad \mbox{if} \; & \theta^0(x,y)& \in (-\frac{\pi}{3},0).
\end{aligned}
\right. $$ 

Denoting  by $\epsilon^{eng}$ the engineering deformation  
$\epsilon^{eng}(t)=(H_0-H(t))/H_0,$
the time period  $[0,T]$   and the strain rate  $d^*=\dot{\Gamma}_c$ were chosen  to correspond to  a final engineering  deformation  $\epsilon^{eng}_{final}=(H_0-H_{final})/H_0=0.5=50\%$ (here $H(t)$ is the actual height of the poly-crystal).

\begin{figure}
	\center
%	\makebox[\textwidth][c]{\includegraphics[scale=0.3]
\includegraphics[height=5cm, angle=0]{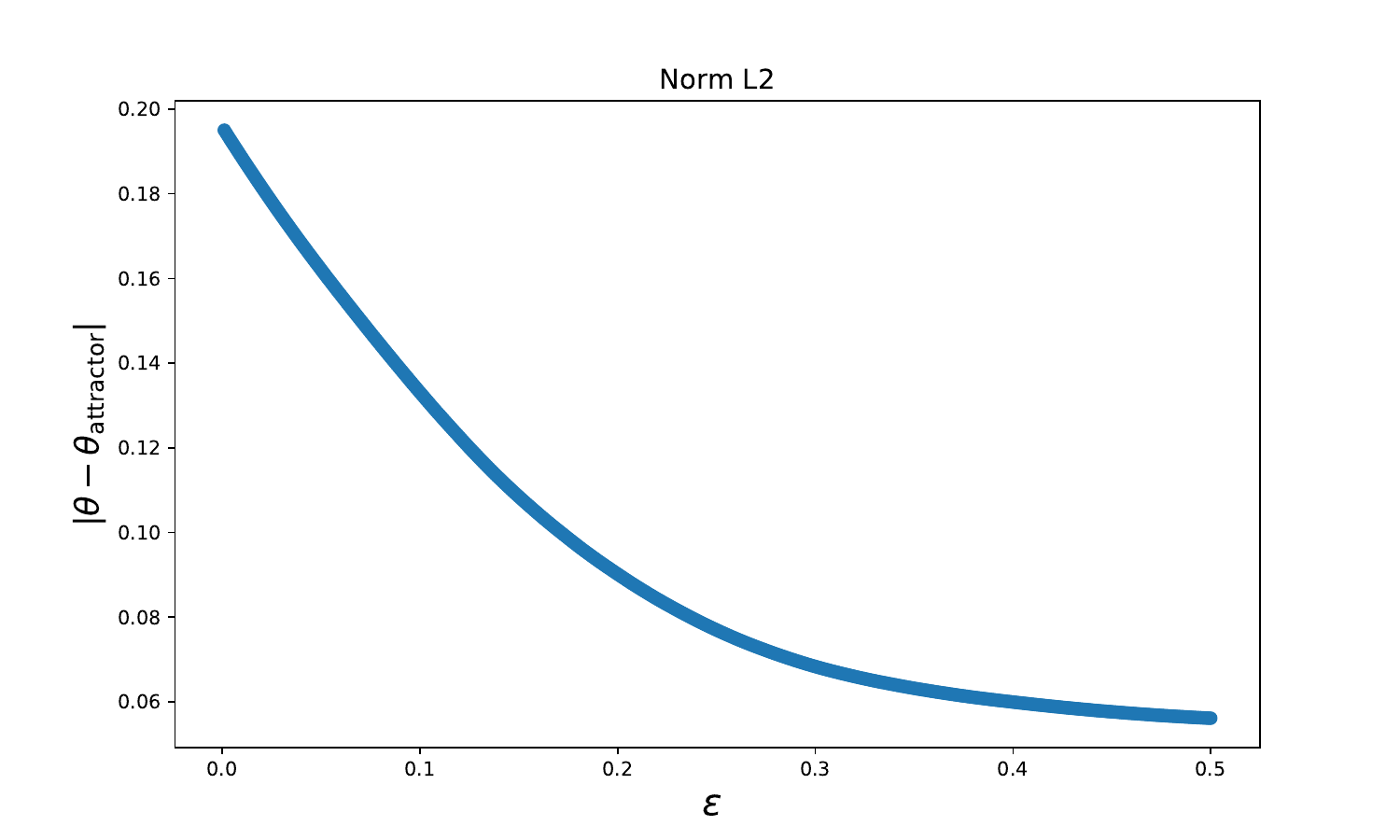}
	\caption{Evolution of the $L^2$  norm  $\vert\vert   g(t) \vert\vert_2$  of the gap between the lattice orientation and the estimated attractor   as a function engineering  deformation  $\epsilon^{eng}$.}
	\label{fig:PolyL2}
\end{figure} 

In Fig.~\ref{fig:PolyOrient} (top) we have plotted the evolution of lattice orientation $(x_1,x_2)\to \theta(t,x_1,x_2)$ during the deformation process. We observe that globally the lattice orientation approaches the estimated attractor $(x_1,x_2)\to \theta^{\text{att}}(x_1,x_2)$ (plotted in Fig.~\ref{fig:PolyOrient} middle) for large strains. The final orientation does not match perfectly with the predicted one but remains very close (less than $5^{\circ}$ almost everywhere). This can be quantified by examining the gap between the lattice orientation and the estimated attractor:
$ (x_1,x_2)\to g(t, x_1,x_2)=\vert \theta(t,x_1,x_2)- \theta^{\text{att}}(x_1,x_2)\vert,$
plotted in Fig.~\ref{fig:PolyOrient} (bottom). To establish a global quantitative measure, we have computed the $L^2$ norm of the gap $g(t)$, given by 
$ \vert\vert g(t) \vert\vert_2=\sqrt{\int_{\Omega(t)} g^2(t, x_1,x_2)\; dx_1dx_2/\text{mes}(\Omega(t))},$
and plotted it in Fig.~\ref{fig:PolyL2}. We observe that the $L^2$ norm shows a significant decrease until $\epsilon^{\text{eng}}=0.3$, after which it exhibits a plateau around $0.06~\text{rad}\approx 3.5^{\circ}$. The small deviations in certain grains can be explained by the loss of strain rate homogeneity and  the presence of shear bands, which initially follow grain boundaries but later appear to be associated with the overall geometry of the sample.

%In Fig.~\ref{fig:PolyOrient} top we have plotted the evolution of lattice orientation  $(x,y)\to \theta(t,x,y)$  during the deformation process. We remark that  globally the lattice orientation is  approaching to the estimated attractor   $(x,y)\to \theta(t,x,y)$  (plotted in Fig. \ref{fig:PolyOrient} middle) for large strains.  The final orientation does not match perfectly with the  predicted one but it is very close (less than $5^\circ$ almost  everywhere). That can be seen from the  gap between lattice  orientation and the estimated attractor  
%$$ (x_1,x_2)\to  g(t, x_1,x_2)=\vert  \theta(t,x_1,x_2)-  \theta^{att}(x_1,x_2)\vert,$$
%plotted in Fig. \ref{fig:PolyOrient} bottom.  To have a global quantitative measure we have computed the $L^2$ norm of the  gap	$g(t)$  given by 
%$$  \vert\vert   g(t) \vert\vert_2=\sqrt{\frac{\int_{\Omega(t)} g^2(t, x_1,x_2)\; dx_1dx_2}{mes(\Omega(t))} },$$
%and plotted  it in Fig.~\ref{fig:PolyL2}. We remark that the $L^2$  norm  has in important decrease  till   $\epsilon^{eng}=0.3$, and then it  exhibits a plateau around    $0.06$rad$\approx 3.5^\circ$.  Only  some  grains  are not very  well estimated. This can be explained by the loss of the strain rate homogeneity, as can be see in Fig.  \ref{fig:PolyMdV}. 
%We remark the presence of some shear bands, which at the beginning follows  grain boundaries but at the end it seems to be associated on the  overall geometry of the sample. 

\begin{figure}	
	\center
%	\makebox[\textwidth][c]{\includegraphics[width=1.1\textwidth]
		\includegraphics[height=4cm, angle=0]{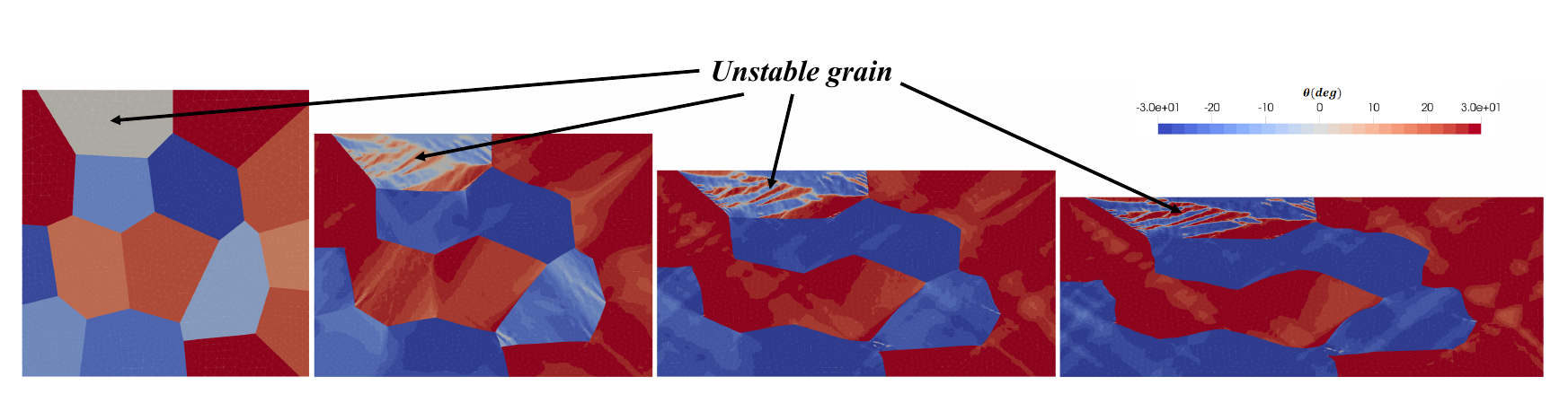}
	\caption{%\textcolor{blue}{
			Computed Eulerian distribution $(x, y) \mapsto \theta(t, x, y)$ of the lattice orientation, shown in a color scale ranging from $-30^\circ$ to $30^\circ$, for the following times: $t = 0$ ($\epsilon^{\text{eng}} = 0$), $t = T/3$ ($\epsilon^{\text{eng}} = 0.166$), $t = 2T/3$ ($\epsilon^{\text{eng}} = 0.33$), and $t = T$ ($\epsilon^{\text{eng}} = 0.5$).
Note that one grain has an initial unstable orientation.%}
}
	\label{fig:PolyOrient2}
\end{figure}

The initial choice of grain orientations, shown in Fig. \ref{fig:Poly} (right), includes only grain orientations that belong to an attraction basin. To explore what happens if a grain has an unstable stationary orientation, we changed the orientation of the upper-right grain (see Fig. \ref{fig:PolyOrient2}) to $\theta = 0$, which is an unstable stationary orientation located between two attraction basins, $(-\pi/3, 0)$ and $(0, \pi/3)$. We observe that part of the grain joins the $\theta^{att} = \pi/6$ attractor, another part joins the $\theta^{att} = -\pi/6$ attractor, while a third part of the grain exhibits orientation bands associated with kink or shear bands. It is important to note that computations in this unstable regime may be mesh or time-step dependent, making it difficult to predict the final configuration.

%The initial choice of  grains orientations, presented in  Fig. \ref{fig:Poly} right, include only  grain  orientations  belonging to an attraction basin. To see what happens if a grain has an unstable stationary orientation we have changed the orientation of upper right grain (see  Fig. \ref{fig:PolyOrient2}) to be $\theta=0$, which is an unstable stationary  orientation located between  two attraction basins $(-\pi/3,0)$ and $(0,\pi/3)$.   We remark that a part of the grain joints $\theta^{att}=\pi/6$ attractor, another part  joints $\theta^{att}=-\pi/6$ attractor, while a third part of the grain presents orientations bands associated to kink or shear bands. We have to mention here that the computations in this unstable regime could be  mesh or time-step dependent, hence it is very difficult to predict the final configuration. 
\subsection{Mono-crystal response to heterogeneous loading} 

\begin{figure}
	\center
%	\makebox[\textwidth][c]{\includegraphics[width=1.\textwidth]
		\includegraphics[height=8cm, angle=0]{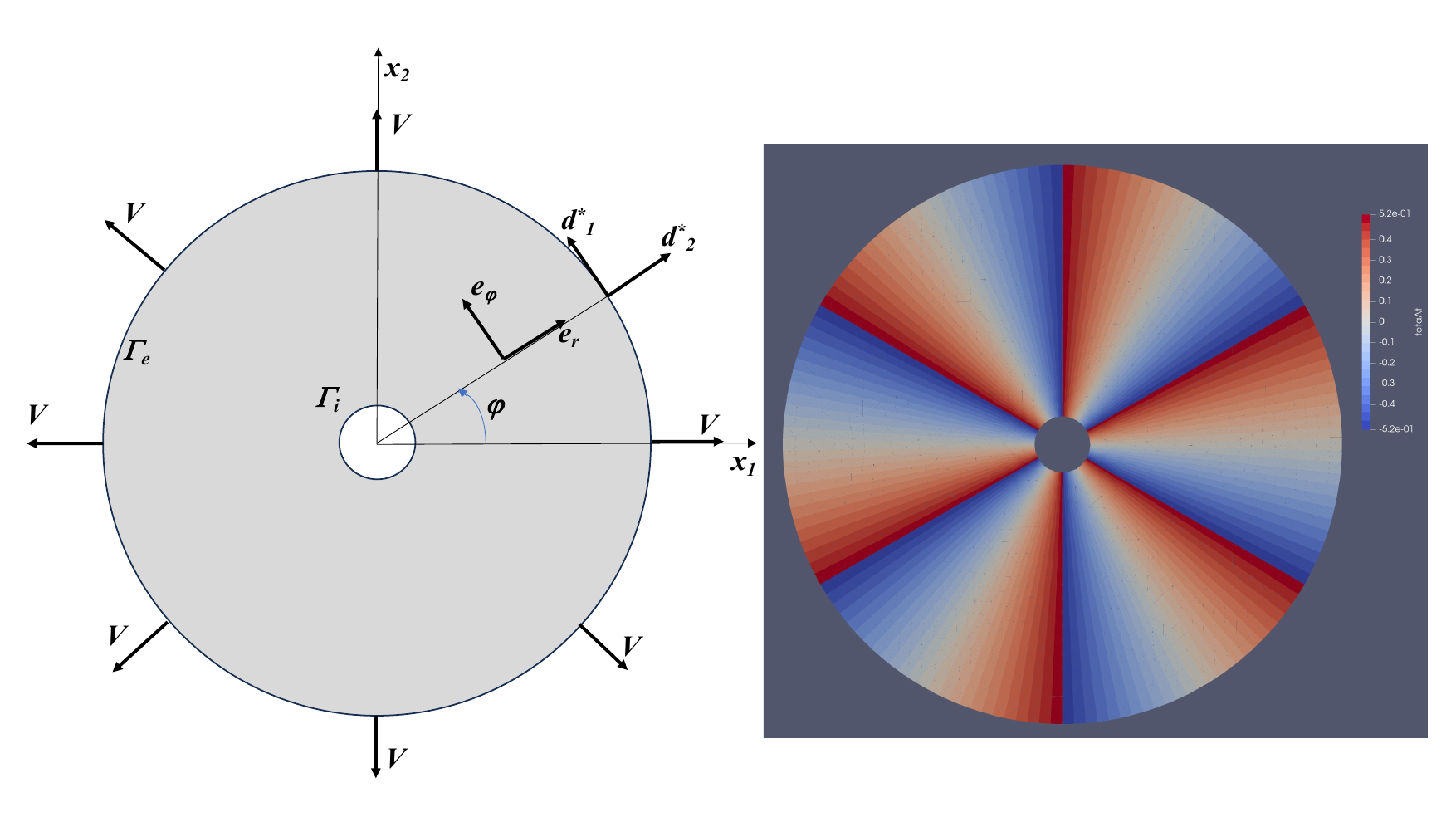}
	\caption{Left: Schematic representation of a single crystal with a circular void at its center, subjected to radial velocity loading.
Right: Distribution $(x, y) \mapsto \theta^{\text{att}}(x, y)$ of the predicted attractor projected on  the initial configuration (radian color scale ranging from $-\pi/6$ to $\pi/6$).}
	\label{fig:Trou}
\end{figure}

At the initial state, the crystal domain is represented by a disc of radius $R_0$, containing an initially circular void at its center with radius $r_0 = R_0/10$ (see Fig.~\ref{fig:Trou}, left). On the external boundary, denoted by $\Gamma_e(t)$, we impose a radial velocity 
\begin{equation}
	{\bV}(t)(x_1,x_2) = \frac{V_0 R_0}{x_1^2 + x_2^2} (x_1 \be_1 + x_2 \be_2),
\end{equation}
corresponding to an incompressible Eulerian velocity field $\bu^*=V_0R_0/r\be_r$, where $r$ and $\varphi$ are the cylindrical coordinates. The internal boundary, denoted by $\Gamma_i(t)$, is traction-free (i.e., ${\Sb}(t) = 0$).
%%%%%%%%%%%%%%%%%%%%%%%%%
%%%%%%%%%%%%%%%%%%%%%%%%%%%%%%%%%%%%%%%%%%%%%%%% 	
The associated gradient of the velocity field $\bu^*$  is $\Lb^*=\D^*=d^*(r)(\be_\varphi\otimes\be_\varphi -\be_r\otimes\be_r )$, with $d^*(r)=V_0R_0/r^2$,
which is non-homogeneous and corresponds to the  choice 
$\bd_1^*=\be_\varphi(\varphi), \quad \bd_2^*=\be_r(\varphi), \quad \omega^*=0, \quad \psi^*=\psi_0^*=\varphi+\frac{\pi}{2}$.

The initial conditions for the velocity and crystal orientation were chosen as
$\bu^0(x_1,x_2) = \bu^*(x_1,x_2), \; \theta^0 = 0$, while 
the time period $[0, T]$, the initial external radius $R_0$, and the velocity $V_0$ were selected to yield a final engineering volumetric strain of $	\epsilon^{eng}_{final} = 2V_0 T/R_0= 0.1875 = 18.75\%.$
Note that the expected deformation at the void level is approximately $R_0 / r_0 = 10$ times higher (around $180\%$) than at the external boundary.

Following the stability analysis of the  rotated orientation angle $\tilde{\theta}=\theta-\psi_0^*$ we deduce that the orientation attractor $\theta^{att}=\tilde{\theta}^{att}+\psi_0^*$  is given on each  of 6 circular sectors of the plane (see Fig. \ref{fig:Trou} right) by 
\begin{equation}\theta^{att}(\varphi) =\varphi -\frac{(k+1)\pi}{3},  \quad \mbox{for}\;  \varphi\in \left(\frac{(1+2k)\pi}{6}, \frac{(3+2k)\pi}{6}\right), \; k\geq 0.\end{equation}

\begin{figure}
	\center
%	\makebox[\textwidth][c]{\includegraphics[width=1.\textwidth]
		\includegraphics[height=7cm, angle=0]{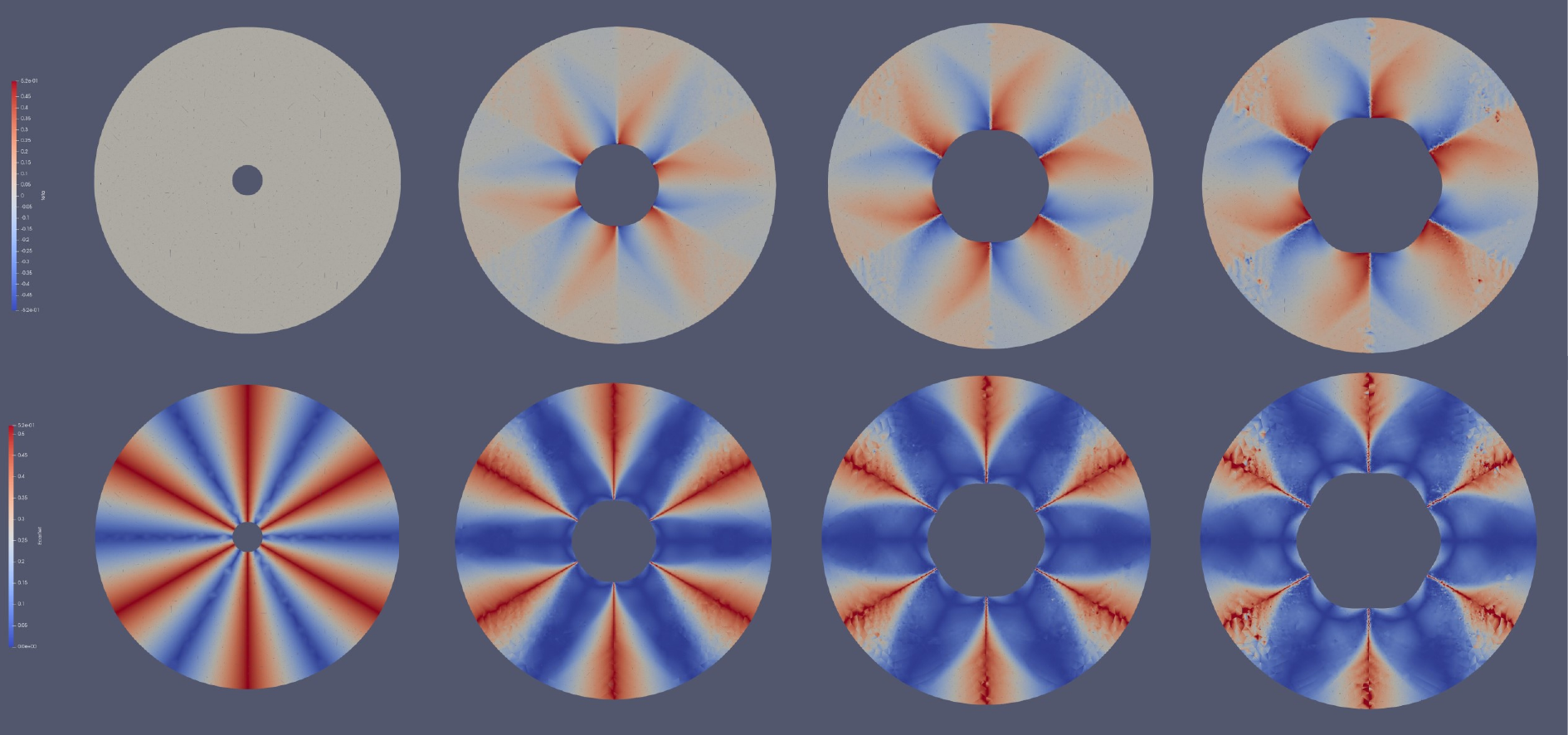}
	\caption{\textcolor{black}{Top: Computed Eulerian distribution $(x, y) \mapsto \theta(t, x, y)$ of the lattice orientation, shown in a radian color scale ranging from $-\pi/6$ to $\pi/6$.
Bottom: Eulerian distribution $(x, y) \mapsto \vert \theta(t, x, y) - \theta^{\text{att}}(t, x, y) \vert$ of the gap between the lattice orientation and the estimated attractor, shown in a radian color scale ranging from $0$ to $\pi/6$.
The distributions correspond to the following times: $t = 0$ ($\epsilon^{\text{eng}} = 0$), $t = T/3$ ($\epsilon^{\text{eng}} = 0.0625$), $t = 2T/3$ ($\epsilon^{\text{eng}} = 0.125$), and $t = T$ ($\epsilon^{\text{eng}} = 0.1875$).}}
	\label{fig:TouOrient}
\end{figure} 

%  	\begin{figure}
%	\center
%%	\makebox[\textwidth][c]{\includegraphics[scale=0.3]
%		\includegraphics[height=5.cm, angle=0]{Chap2/Images/RadialMdV.pdf}
%	\caption{The computed equivalent strain rate for  $t=T/3$ ($\epsilon^{eng}=0.0625$), $t=2T/3$  ($\epsilon^{eng}=0.125$) and $t=T$  ($\epsilon^{eng}=0.1875$).}
%	\label{fig:TouMdV}
%\end{figure} 

In Fig.~\ref{fig:TouOrient} (top), we plot the evolution of the lattice orientation $(x,y) \mapsto \theta(t,x,y)$ during the deformation process. We observe that the orientation exhibits discontinuities at the angles $\varphi = \pi/6 + k\pi/3$, consistent with the attractor structure. However, these discontinuities are more pronounced near the void, where deformations are significantly larger, than near the external boundary. This observation aligns with our stability analysis from the previous section: the distance between the lattice orientation and its attractor decreases substantially for large strains (greater than $50\%$), which occur near the void, but remains small at the external boundary.

In Fig.~\ref{fig:TouOrient} (bottom), we show the distribution $(x,y) \mapsto |\theta(t,x,y) - \theta^{att}(t,x,y)|$, representing the difference between the lattice orientation and the estimated attractor. We observe that, overall, the blue regions (where the orientation is close to the attractor) expand during the deformation. However, as discussed above, there are regions—namely along the discontinuity lines $\varphi = \pi/6 + k\pi/3$ and areas far from the void (shown in red)—where the attractor fails to accurately describe the lattice orientation.

\begin{figure}
	\center
%	\makebox[\textwidth][c]{\includegraphics[width=1.\textwidth]		
\includegraphics[height=7cm, angle=0]{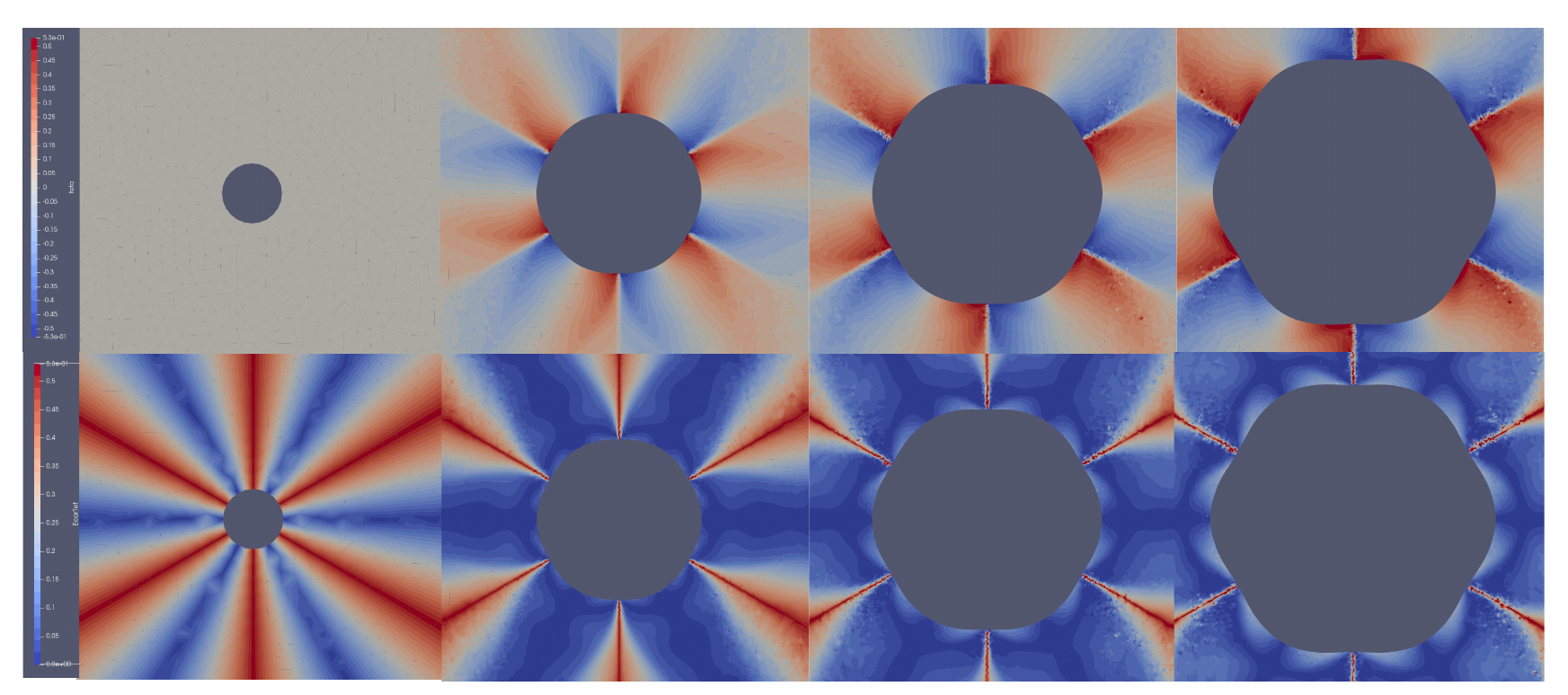}
\caption{%\textcolor{blue}{
		Magnified view of the void region highlighted in Fig. \ref{fig:TouOrient}. Top: Computed Eulerian distribution of lattice orientation $(x,y)\to \theta(t,x,y)$ shown in radian color scale $(-\pi/6,\pi/6)$. Bottom: Eulerian distribution of deviation between lattice orientation and estimated attractor $(x,y)\to \vert\theta(t,x,y)-\theta^{att}(t,x,y)\vert$ displayed in radian color scale $(0,\pi/6)$. Distributions are presented at four deformation stages: $t=0$ ($\epsilon^{eng}=0$), $t=T/3$ ($\epsilon^{eng}=0.0625$), $t=2T/3$ ($\epsilon^{eng}=0.125$) and $t=T$ ($\epsilon^{eng}=0.1875$).%}
	}
	\label{fig:TouOrientZoom}
\end{figure}

We should also mention that very close to the void, at large strains, the deformation process becomes much more complex: slip and kink bands appear  (see a detailed analysis in \cite{Salman2021-jh,Smiri:2024aa,SPSI25}). Moreover, during the deformation process, the computed equivalent strain rate is no longer radial, as expected from $\D^*$. This is why the previous analysis, which relies on a simple assumption about the velocity field $\bu^*$, is less pertinent. This can be observed in Fig. \ref{fig:TouOrientZoom} (bottom), where at $t = 2T/3$ ($\epsilon^{eng} = 0.125$) and at $t = T$ ($\epsilon^{eng} = 0.1875$), the difference between the lattice orientation and the estimated attractor is larger very close to the void. 

\subsection{Slip band deformation  in a mono-crystal} 

\label{SlipBandsSection}

\begin{figure}%[t]
	\center
%	\makebox[\textwidth][c]{
		\includegraphics[height=8cm, angle=0]{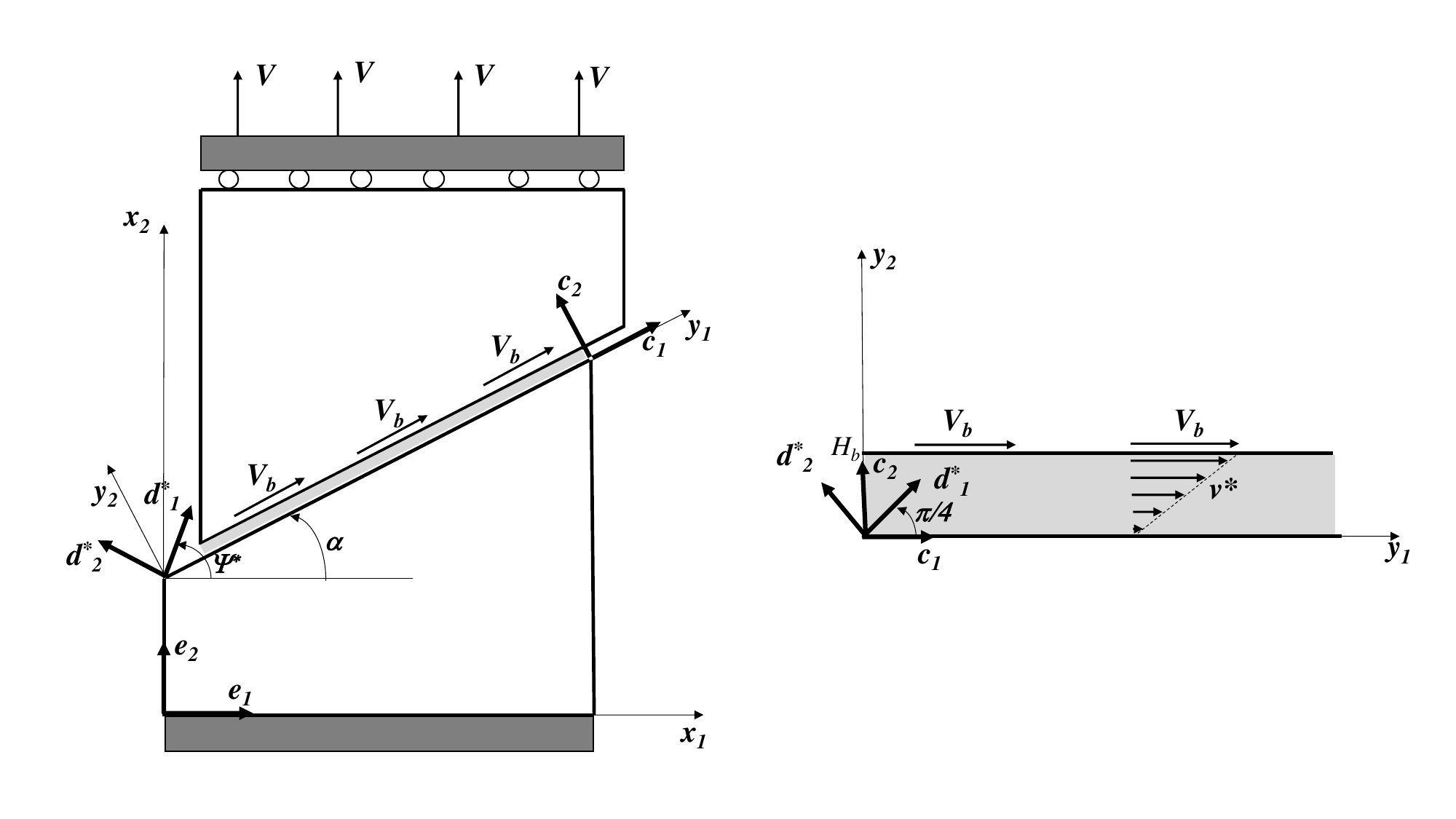}
		\vspace{-0.5cm}
	\caption{%\textcolor{blue}{
			Schematic description of a slip band  under tensile loading.%}
		}
	\label{SlipBand}
\end{figure}

The mono-crystal domain, which   has a  initial rectangular shape $(0,L)\times(0,H)$,  is pulled on the upper  part with a normal velocity $V$ and it is fixed on bottom.  All other boundaries are stress free.  We analyze  one   slip band  localized  deformation of width $H_b$ and orientated at the angle $\alpha$  with respect to the $Ox_1$ axis.  In the slip band  we expect  the velocity  distribution $\displaystyle \bu^*(y_1,y_2)=V_b\frac{y_2}{H_b}\bc_1$, where we have denoted by $\bc_1, \bc_2$ the basis associated to the shear band and by $(y_1,y_2)$ the shear band variables (see Fig. \ref{SlipBand} ), and $V_b=V/\sin(\alpha)$.  Denoting by $2d^*=V_b/H_b$ we deduce that the expected velocity gradient in the shear band  is 
$\Lb^*=2d^*\bc_1\otimes\bc_2,  \quad   \D^*=d^*(\bc_1\otimes\bc_2 +\bc_2\otimes\bc_1 ), \quad  \bW^*=d^*(\bc_1\otimes\bc_2 -\bc_2\otimes\bc_1).$ 
We can compute the principal strain rate directions $\bd_1 ^*=(\bc_1+\bc_2)/\sqrt{2},\bd_2^* =(\bc_2-\bc_1)/\sqrt{2}$  to deduce that  we deal with a stationary velocity gradient with $\psi^*=\alpha+\pi/4$.  
$\bu^0 (x_1,x_2)=V\frac{x_2}{H}\be_2, \quad \theta^0=0$, while 
the time period  $[0,T]$ and the velocity $V$ were chosen to correspond to  a final engineering deformation  $\epsilon^{eng}_{final}=VT/H=0.2=20\%$.    Note that the expected deformation  in the  slip band is expected to be much larger. 

\begin{figure}	
%	\center
%	\makebox[\textwidth][c]{\includegraphics[scale=0.3]
		\includegraphics[height=8cm, angle=0]{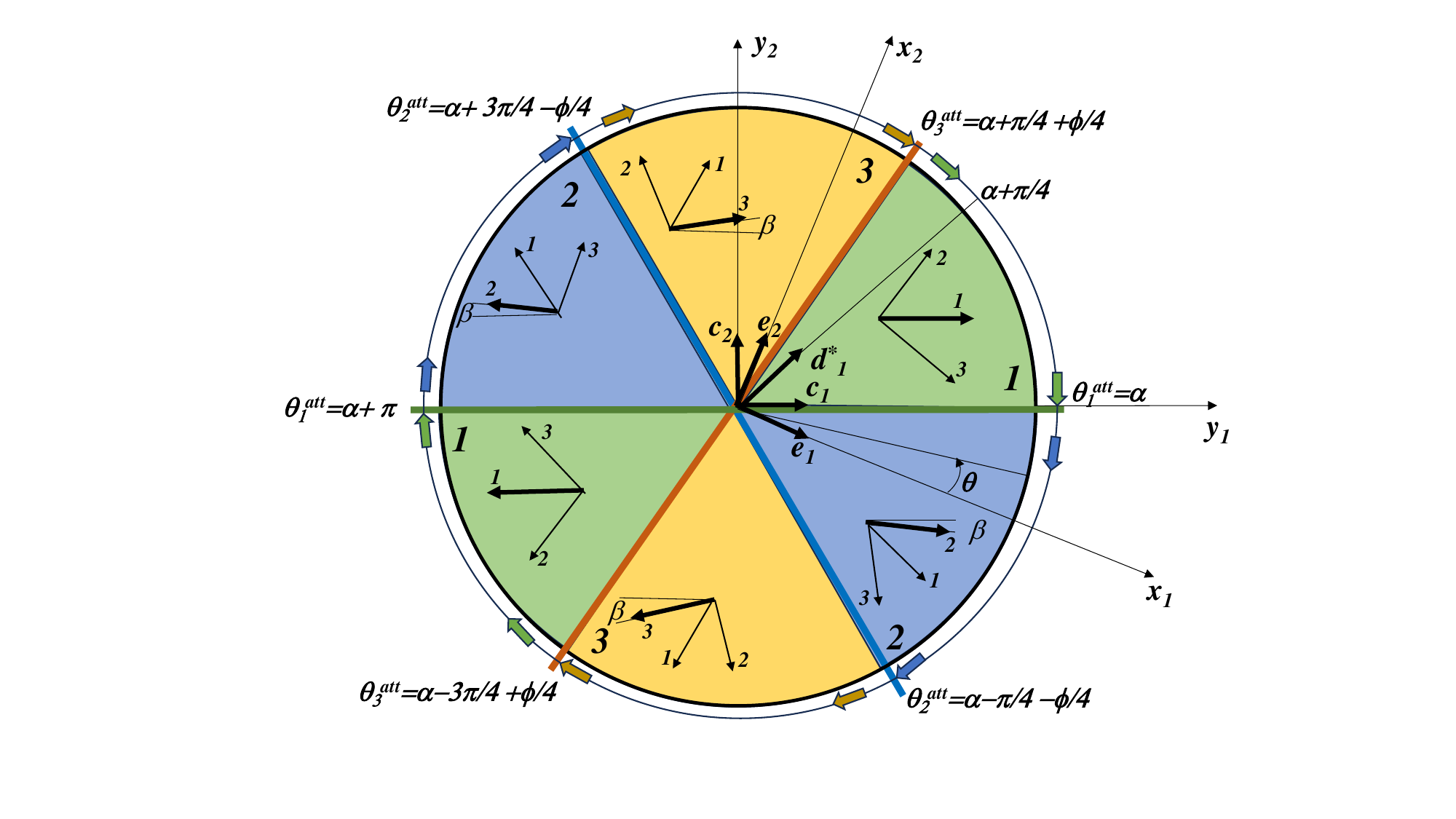}
	\caption{
		%\textcolor{blue}{
			Stability analysis for a slip band under tensile loading ($\alpha>0$): Schematic representation of three half-attractors $\theta_s^{att}$ with their corresponding attraction basins $B^{att}_s$ (green for $s=1$, blue for $s=2$, and orange for $s=3$). For each half-attractor, the single active slip direction is highlighted in bold. The parameter $\beta$ represents the misorientation between the active slip system and the slip band orientation.%}
		}
%	\caption{Stability analysis for a slip band in traction ($\alpha>0$): schematic description of  three  half-attractors  $\theta_s^{att}$  with their attraction basins  $B^{att}_s$  (green for  $s=1$,  blue for $s=2$ and orange for $s=3$). For each  half-attractor  the single active slip direction is in bold while  $\beta$ is the miss-orientation between the active slip system and  slip band orientation.}
	\label{SlipBandAttractor}
\end{figure}

Let use here the stability analysis developed in section \ref{StabAnalStationary} to predict the lattice orientation in a  slip band in traction with  $\alpha<\pi/2$. First, we  remark that we  deal with  case ii) with $\omega^*=d^*$ where there are three  half-attractors (i.e., attractors on the left but not on the right)  $\theta_i^{att}=\tilde{\theta}_i^{att}+\alpha+\pi/4$, where $\tilde{\theta}_i^{att} $  are given by (\ref{Att2p}),  
while the  3 basins of attraction are given by  (\ref{Att2pB})   translated with $\alpha+\pi/4$. 
Note that each attractor corresponds to a single active system and the basins label $s$ have been chosen to correspond to the active slip system $s$ (see Fig. \ref{SlipBandAttractor}).  If we denote by $\beta_s$ the miss-orientation between the active slip system of the attractor $s$ and the slip band direction we get that 
$$\beta_1=0, \quad  \beta_2=\beta_3=\frac{\pi}{4}- \frac{3\phi}{4}.$$
We deduce that  the first attractor  has the active slip  system parallel ($\beta_1=0$) with the slip band.  We arrive at the same conclusion for the other two attractors  ($\beta_2=\beta_3=0$) in the case of a hexagonal crystal ($\phi=\pi/3$). For a FCC  crystal  ($\phi=54.7^\circ$) there exists a slight miss-orientation   $\beta_2=\beta_3\approx 4^\circ$,  and  the stationary slip band  could  be re-considered as a stationary  kink band.

In conclusion, for a slip band under traction, the stability analysis predicts that the lattice orientation of the crystal within the band will rotate in a clockwise direction, such that one of the slip systems will become parallel (or nearly so, in the case of an FCC crystal) to the slip band. It is important to note that in the above analysis, we assumed that the direction of the slip band is given and independent of the lattice rotation. However, this is not generally the case, as a more complex process is expected, where the sample shape, boundary conditions, and crystal orientation interact to explain the existence of one or more slip bands and their orientations.

\begin{figure}%[t]
%	\center
%	\makebox[\textwidth][c]{\includegraphics[scale=0.3]
		\includegraphics[height=5.5cm, angle=0]{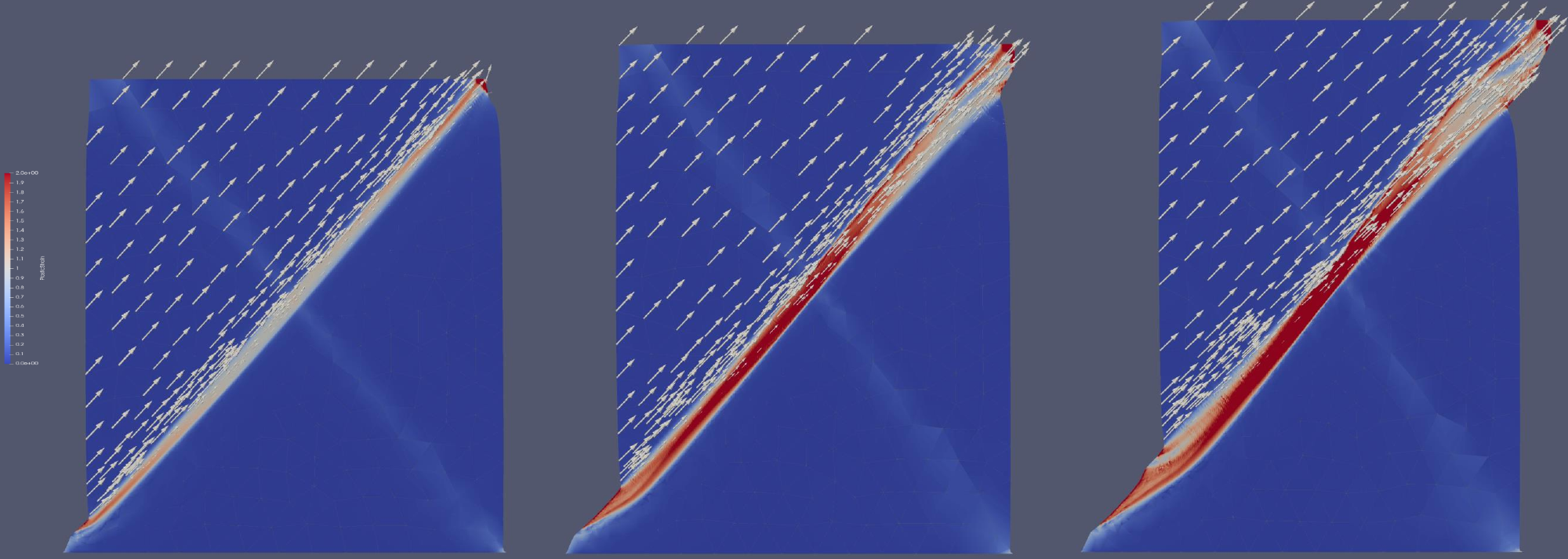}
%	\vspace{-.5cm}
\caption{The computed accumulated plastic strain $\epsilon^p(t)$ (in color scale) and the velocity distribution (arrows) at three deformation stages: $t=T/3$ ($\epsilon^{eng}=0.067$) (left); $t=2T/3$ ($\epsilon^{eng}=0.133$) (center); and $t=T$ ($\epsilon^{eng}=0.2$) (right).}
	\label{StrainSlipBan}
\end{figure}

In Fig. \ref{StrainSlipBan} we have plotted the evolution of the deformation. We remark  the presence of a slip band orientated at $\alpha=\pi/4$ where the plastic strain $\epsilon^p$   (i.e., $\dot{\epsilon}^p = \vert \Db(\bu)\vert$ )   is accumulated during the process. We remark that  the deformation assumption  with two rigid blocks separated by a slip band (see Fig. \ref{SlipBand}) is globally  verified.  Moreover, the level of plastic strain in the slip band is important (more than $2=200\%$)  hence the above stability analysis could be  pertinent.

\begin{figure}%[t]	
	\center
%	\makebox[\textwidth][c]{\includegraphics[scale=0.3]
	\includegraphics[height=8.cm, angle=0]{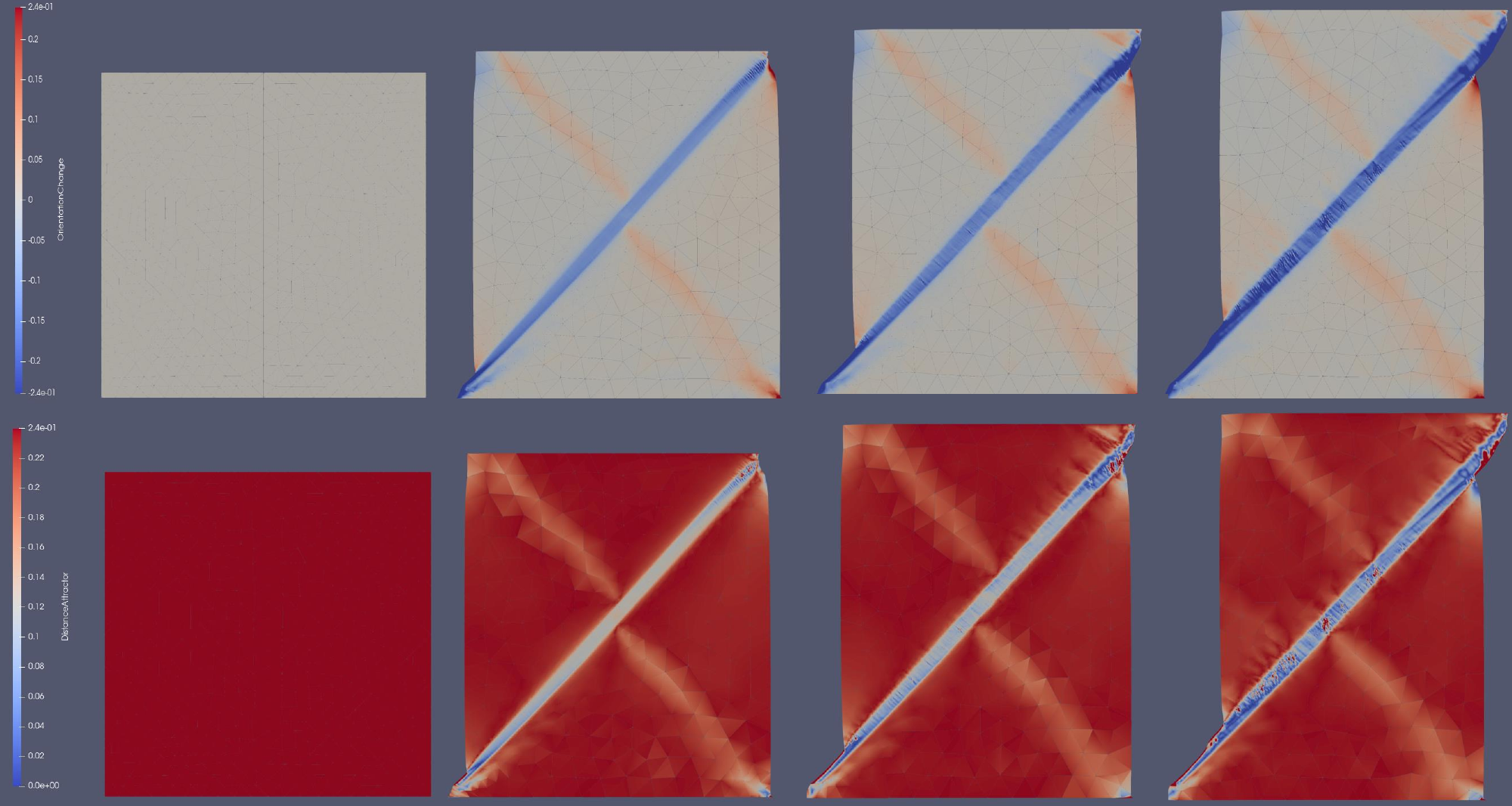}
%	\vspace{-.5cm}
	\caption{Up:  the lattice orientation $\theta(t)$ (in rad  ranging between $-\phi/4$ and $\phi/4$).  Bottom: the  gap $\min_s\vert \theta(t)-\theta_s^{att}\vert$ between the lattice orientation and the slip band  attractor (in rad ranging between $0$ and $\phi/4$).   For $t=0$ ($\epsilon^{eng}=0$), $t=T/3$ ($\epsilon^{eng}=0.067$),  $t=2T/3$ ($\epsilon^{eng}=0.133$)  and   $t=2T/3$ ($\epsilon^{eng}=0.2$).}
	\label{OrientatiionSlipBand}
\end{figure}

As it follows from the above stability analysis (see also Fig. \ref{SlipBandAttractor}) the initial orientation of the crystal $\theta_0=0$  is in  attraction basins $B_2^{att}$ with the attractor $\theta_2^{att}=-\phi/4$ hence  the expected  lattice orientation in the slip band is $-\phi/4\approx-13.7^\circ=-0.24$rad. In Fig. \ref{OrientatiionSlipBand} we have plotted the evolution of the distribution of the lattice orientation $\theta(t)$ (in rad, up) and of the its gap  $\min_s\vert \theta(t)-\theta_s^{att}\vert$  with  the slip band  attractor (in rad, bottom). We remark that, as expected,  the orientation in  the slip band changes during the slip to reach asymptotically the  attractor.

\begin{figure}[t]	
	\center
%	\makebox[\textwidth][c]{\includegraphics[scale=0.3]
		\includegraphics[height=6cm, angle=0]	{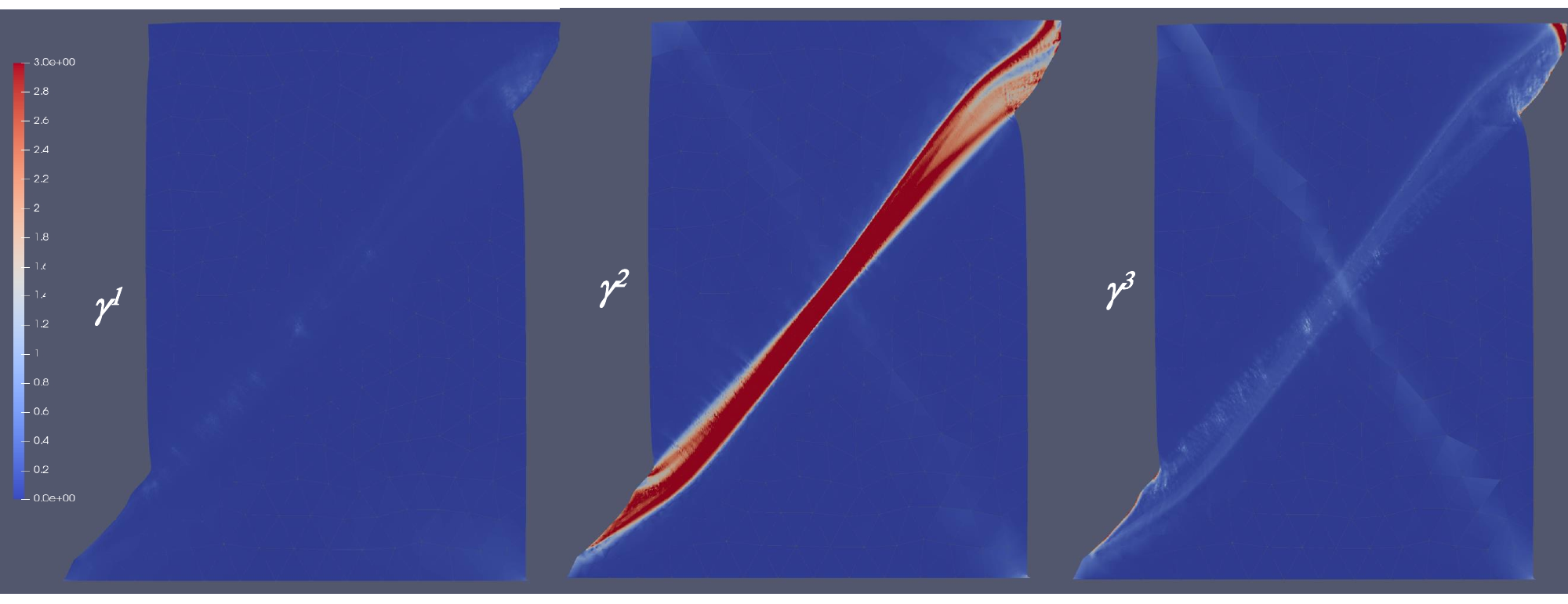}
\caption{%\textcolor{blue}{
		The computed distribution of the accumulated plastic slip $\gamma^s_a$ ($\dot{\gamma}_a^s=\vert\dot{\gamma}^s\vert$) at the end of the process $t=T$ ($\epsilon^{eng}=0.2$) for slip systems: $s=1$ (left), $s=2$ (center), and $s=3$ (right).%}
	}	\label{SlipFinal}
\end{figure}

Concerning the active slip system in the slip band we expect that  only system $s=2$  is active in the slip band. To check this we have plotted in Fig. \ref{SlipFinal} the computed distribution of the  accumulated plastic slip  $\gamma^s_a$  (i.e., $\dot{\gamma}_a^s=\vert\dot{\gamma}^s\vert$) at the end of the process  $t=T$ ($\epsilon^{eng}=0.2$). We remark that in the shear band only the system $s=2$ is active. 

%%%%%%%%%%%%%%%%%%%%%%%%%%%%%%%%%%%%%%%%%
\section {Conclusions}

In this paper, we focused on the stability of the lattice orientation for velocity gradient driven processes. To simplify the problem, we considered only rigid visco-plastic models, without hardening/softening effects. We formulated the general problem of lattice orientation stability analysis explicitly in a mathematical framework where tools such as linear stability could be used. This analysis allowed us to select the attractors from among the stationary lattice orientations and hence to predict the final texture of the poly/monocrystal for a given velocity gradient driven process.

However, we found that applying this stability analysis to a specific 3-D crystal (FCC, HCP, etc.) is very complex due to two main difficulties: the analytic description of the slip rates as functions of lattice orientation and the non-linearity of the problem. The detailed analysis of the 3-D case, which is beyond the scope of our present paper, could be considered in future work. In this study, we restrict our analysis to a simplified 2D model incorporating
three slip systems. Within this framework, we develop closed-form analytical expressions for
the slip rates as functions of both lattice orientation and strain rate. We used this key result in our linear stability analysis to characterize the stable stationary orientations and to compute their attractors. Since the attractors are effective (i.e., the distance between the lattice orientation and the attractor is small) only for slip strains larger than 50\%, this stability analysis, which concerns large deformations, is not so useful for small strains.

To illustrate our stability analysis, we examined three specific problems in CP.  In all cases, we compared the numerical results of the associated boundary problem with the theoretical prediction. For numerical integration, only the boundary conditions with respect to the driven velocity field are provided. This results in a computed velocity field that differs from the expected one.  However, for the final overall orientation, we found good agreement between the predicted attractor and the numerical results.

In the first case, we considered a homogeneous driven velocity gradient field with non-homogeneous initial orientations, modeling the grains of a polycrystal. Since we initialized all grains with orientations belonging to one of two attractors, our stability analysis yielded the expected final texture of the polycrystal. We found the computed orientation to be in very good agreement with our predicted one, with a gap (in $L^2$ norm) of less than $4^\circ$. We also analyzed the situation when only one grain had an initial unstable orientation. We observed that this grain developed many shear bands, but globally it had a limited effect on the global orientation of the polycrystal. This comparison illustrated how subtle variations in initial crystallographic orientations can significantly affect the distribution of strain within individual grains.

In the second case, we studied a mono-crystal (homogeneous initial orientation) with a non-homogeneous driven velocity gradient: void growth under radial loading. Our numerical simulations showed that, in accordance with our stability analysis, the orientation at large strains exhibited a discontinuity at the angles $\varphi=\pi/6+k\pi/3$, as the attractor does. We observed these discontinuities near the void, where the deformations were large enough, and they vanished near the outer boundary, where the deformation was too small for the attractor to be effective. Very close to the void, at large strains, we found that the deformation process was much more complex: slip and buckling bands appeared and the strain rate differed significantly from the driven one. Therefore, we measured a larger difference between the lattice orientation and the estimated attractor very close to the void.

The third case we investigated was a slip band in uni-axial traction of a mono-crystal. Our stability analysis of the slip band predicted that the lattice orientation of the crystal in the band would rotate such that one slip system would become parallel with the slip band. This slip system is the only stationary active slip system in the slip band. However, there are some configurations associated with  in-plane FCC crystals,  where  a small  miss-orientation ($4^\circ$) between the slip direction and the slip band could exists. In our theoretical analysis, we supposed that the slip band orientation is  known and  fixed. But,  in applications, there exists a complicated interplay between the sample shape, the boundary conditions, and the initial crystal orientation which generates the existence of one or several slip bands and their orientations.

We designed our numerical simulations to illustrate the assumption of two rigid blocks separated by one slip band with a high level of localized plastic strain (more than 200\%) such that our stability analysis could be pertinent. As we expected, the orientation in the slip band changed during the slip to reach asymptotically the attractor with only one active slip system.

Finally, our theoretical stability analysis, validated through numerical simulations, provides
a comprehensive framework for understanding how initial crystallographic orientations, deformation mechanisms, and applied loading conditions govern lattice orientation evolution.
Moreover, our theoretical characterization of orientation attractors enables reliable prediction
of crystal texture development under large deformations.

{\bf Acknowledgments.}  I. R. I. thanks Prof. Armand Beaudoin for interesting and useful discussions, which generated his interest in this subject. O. U. S. was supported by grants ANR-18-CE42-0017-03, ANR-19-CE08-0010-01, and ANR-20-CE91-0010.

%\bibliographystyle{abbrv}
%\bibliography{bib1_formatted-1}

\end{document}